\documentstyle[aps,prb,preprint,amsfonts]{revtex}
\begin{document}
\draft
\title{Two-dimensional electron transport\\
in the presence of magnetic flux vortices}
\author{Mads Nielsen\cite{adress} and Per Hedeg\aa{}rd\cite{adress}}
\address{\O{}rsted Laboratory, Niels Bohr Institute, Universitetsparken 5,
2100 Copenhagen \O, Denmark.}
\date{\today}
\maketitle

\begin{abstract}
We have considered the conductivity properties of a two dimensional electron
gas (2DEG) in two different kinds of inhomogeneous magnetic fields, i.e.\
a disordered distribution of magnetic flux vortices, and a periodic array of
magnetic flux vortices. The work falls in two parts. In the first part
we show how the phase shifts for an electron
scattering on an isolated vortex, can be calculated analytically, and related
to the transport properties through a force balance equation. In
the second part we present numerical results for the Hall conductivity
of the 2DEG in a periodic array of flux vortices.
We find characteristic peaks in the Hall
conductance, when plotted against the filling fraction.
It is argued that the peaks can
be interpreted in terms of ``topological charge'' piling up across local and
global gaps in the energy spectrum.
\end{abstract}

\section{Introduction}
Over the last decade the two dimensional electron gas (2DEG) have been
exposed to a wide range of physical experiments, in which the electrons have
been perturbed by different configurations of electrostatic potentials, with
or without a homogeneous perpendicular magnetic field. These experiments
have shown new kinds of oscillations in the magnetoconductivity, with a
periodicity not given by the geometry of the Fermi surface,
as is the case with the Shubnikov-de Haas oscillations,
but given by the interaction of the two
length scales given respectively by the magnetic length, and by the
spatial structure of the potential, e.g.\ the Weiss
oscillations~\cite{WeissOsc}.
More recently, there have been increasing interest in systems
where the 2DEG is exposed to an {\em inhomogeneous}
perpendicular magnetic field. In such systems the inhomogeneities in the
magnetic field acts as perturbations of the 2DEG, relative to the homogeneous
magnetic field, where the band structure consists of the dispersionless
Landau bands. The inhomogeneous magnetic field appears in the Hamiltonian
in the form of a non trivial vector potential. In the case of a periodic
variation in the magnetic field, it is possible to construct a periodic
vector potential, if and only if the flux through the unit cell of the field
is equal to a rational number, when measured in units of the flux quantum
$\phi_0=h/e$. Under these special circumstances
Bloch states can be used as a basis for the calculation of response
properties of the electron gas.

In this paper we have considered a special class of spatially varying
magnetic fields which consists of flux vortices, that are either distributed
at random or placed in a regular lattice structure.
A system consisting of a 2DEG penetrated by a random distribution
of magnetic flux vortices, have been
experimentally realized by Geim et al.~\cite{Geim92,Geim93}.
They made a sandwich construction of
a GaAs/GaAlAs sample with a 2DEG at the interface, and a type~II
superconducting lead film (electrically disconnected from the 2DEG).
When the system was placed in an
external magnetic field, and cooled below the transition temperature of the
film, the magnetic field penetrated the film, and thereby the 2DEG, in the
form of Abrikosov vortices. When the external magnetic field is weak,
below 100G, the vortices will be well separated, and the 2DEG therefore sees
a very inhomogeneous magnetic field. In the experiments conducted by Geim
et al.\ the flux pinning in the film was strong, resulting in a disordered
distribution of flux vortices. This is the physical situation which we
investigate in Sec.~\ref{sec:single} below. In very clean films of type~II
superconducting material, the flux vortices will order in a periodic array,
i.e.\ an Abrikosov lattice, and thereby create a periodic magnetic field
at the 2DEG. This is the situation which we analyse in Sec.~\ref{sec:array}.

Several authors have investigated the transport properties of 2DEG's in
different kinds of inhomogeneous magnetic fields. Peeters and
Vasilopoulos~\cite{Peeters} have made a theoretical study of
the magnetoconductivity in a 2DEG in the presence of a magnetic field,
which was modulated weakly and periodically along one direction. They found
large oscillations in the longitudinal resistivity as a function of the
applied magnetic field strength. These oscillations are due to the
interference between the two length scales given respectively by the period
of the lateral variation of the magnetic field, and by the magnetic length
corresponding to the average background field. The oscillations are
reminiscent of the Weiss oscillations, but have a higher amplitude and a
shifted phase, relative to the magnetoresistance oscillations induced by the
periodic electrostatic potential.

The problem of how the transport properties of the 2DEG is modified by the
presence of a random distribution of flux vortices, have been treated earlier
by A. V. Khaetskii~\cite{Khaetskii91}, and also by
Brey and Fertig~\cite{Brey}.
The approach used by these authors are closely related to the one we have
presented in Sec.~\ref{sec:single}, i.e.\ based on the scattering theory,
and the Boltzmann transport equation.
Khaetskii considered the scattering in certain limiting cases,
including the semiclassical, and Brey
and Fertig calculated the transport properties numerically.

In Sec.~\ref{sec:array} we will address the ``paradox'' of how the Hall effect
can disappear in the following situation: We imagine a 2DEG in a regular
2D-lattice of flux vortices, with the magnetic field from a single vortex
exponentially damped with an exponential length $\xi$, in units of the lattice
spacing. We take the total flux from a single vortex to be $\phi_0/2$, as is
the case when the vortices come from a superconductor. When $\xi\gg 1$, the
field is homogeneous and the Hall conductivity is $\sigma_H =p e^2 /h$,
where $p$ is the number of filled bands. In the other limit i.e.\ when
$\xi\rightarrow 0$, the time reversal symmetry is restored, and the Hall
conductivity vanish. The fact that the system has time reversal symmetry
when the vortices are infinitely thin, can be seen by subtracting a Dirac
string carrying one quantum of magnetic flux $\phi_0$, from each flux vortex.
The introduction of the Dirac strings can not change any physical quantities,
and the procedure therefore establishes that the system, with infinitely thin
vortices with flux $+\phi_0/2$, is equivalent with the system with reversed
flux $-\phi_0/2$, through each vortex. The paradoxical situation arises
because it is known from general arguments~\cite{TKNN82}, that the
contribution to the total Hall conductivity from a single filled
nondegenerate band is a topological invariant, and therefore cannot change
gradually. The situation is even more clear cut if we imagine a periodic
array of exponential flux vortices carrying one flux quantum $\phi_0$ each.
Then the limit $\xi=\infty$ corresponds to a homogeneous magnetic field,
while the opposite limit $\xi=0$ coresponds to free particles.
Incidentally this scheme can be used to establish an interpolating path
between the multi-fractal structure known as Hofstadters
butterfly~\cite{Hofstadter} which is a plot of the allowed energy levels for
electrons on a lattice in a homogeneous magnetic field, and the corresponding
plot for lattice-electrons in no field,
which is completely smooth.

The plan of the paper is as follows.
In Sec.~\ref{sec:single} we calculate the longitudinal and transverse
resistivities of the 2DEG, in the limit where the mean distance between the
vortices is so long that correlation effects are negligible.
First we review the classical scattering theory in Sec.~\ref{sec:class},
before we discuss the quantum dynamics in Sec.~\ref{sec:quantum}.
The longitudinal and transverse resistivities are discussed in
Sec.~\ref{sec:moment}, and resonance scattering is demonstrated
in Sec.~\ref{sec:reso}.
The case of a 2DEG in a periodic array of flux
vortices, is the subject of Sec.~\ref{sec:array}. In the first part,
Sec.~\ref{sec:intro}--\ref{sec:bandcross}, of this
section the general theory of electron motion in a periodic magnetic field
is reviewed, and in the second part, Sec.~\ref{sec:num}--\ref{sec:sup},
we present the results of the numerical calculations.

\section{Single Vortex Scattering}\label{sec:single}

In this section we will consider the introduction of
magnetic flux vortices into a 2 dimensional electron gas,
in the limit where each vortex can be  treated as an individual
scattering center. The vortices
are assumed to be distributed at random, homogeneously over the sample. The
average separation between the vortices is assumed so large, that we can
neglect interference from multiple scattering events.

\subsection{The Classical Cross Section}\label{sec:class}

We will start by calculating the differential cross section for a charged
particle
scattering on a flux vortex within the framework of classical mechanics. This
will provide a reference frame, and allow us to speak unambiguously about the
classical limit. In the calculations we will use an ideal vortex, which has
a circular cross section with constant magnetic field inside, and zero
magnetic field outside
\begin{equation}\label{eq:bfield}
B(\bbox{r})=\left\{ \begin{array}{ccc}
\displaystyle{\frac{\phi}{\pi r_0^2}} & \mbox{for} &  r<r_0 \\
0                          & \mbox{for} &  r>r_0 .
\end{array}\right.
\end{equation}
Here $r_0$ is the radius, and $\phi$ is the total flux carried by the vortex.
The classical orbit is found as the
solution to Newton's equation of motion with the force given by the Lorentz
expression $\bbox{F}=e\bbox{v\times B}$, for a particle of charge $e$.
It consists, as is well known, of
straight line segments outside the vortex, and an arc of a circle inside, with
radius of curvature given by the cyclotron radius $l_c =v/\omega_c$, with
$v$ being the particle velocity, and $\omega_c =\frac{eB}{m}$ the cyclotron
frequency. Because the orbit inside the vortex is an arc of a circle, and it
is impossible to draw a circle that only cut the circumference of the vortex
once, it follows that a particle which initially is outside the vortex,
can never become trapped in the vortex.
The geometry of the classical scattering process shown in Fig.~\ref{fig:geo},
is determined by the dimensionless parameter $\gamma=l_c/r_0$, which is the
ratio between the radius of the cyclotron orbit,
and the radius of the flux vortex.
The definitions of the reduced impact parameter $\beta=b/r_0$, and the angles
$\phi,\psi,\theta$, follows from Fig.~\ref{fig:geo}.
By inspecting the figure, it is observed that the following relations hold
\begin{eqnarray}
  \beta  & = & \sin \phi  \\
  \tan \psi & = & \frac{\gamma+\beta}{\sqrt{1-\beta^2}} \\
  \gamma\sin\frac{\theta}{2} & = & \sin (\psi - \phi)
  \text{sign}(\gamma+\beta),
\end{eqnarray}
where the sign of $\gamma$ is dictated by the direction of the magnetic field
inside the vortex. Hereafter we take $\gamma$ to be positive. After a small
amount of arithmetic
$\phi$ and $\psi$ are eliminated, and we have
\begin{equation}\label{eq:scatt}
\sin\frac{\theta}{2} = \text{sign}(\gamma+\beta)
\sqrt{\frac{1-\beta^2}{\gamma^2 +2\gamma\beta +1}}.
\end{equation}
This relation gives the scattering angle $\theta=\theta(\beta)$ as a function
of the impact parameter $\beta$. It is observed that for $0<\gamma<1$ the
scattering angle $\theta$ sweeps through the interval $[-\pi,\pi ]$, for
$\beta\in [-1,1]$, while for $\gamma>1$ we have $\theta\in [0,\theta_0 ]$,
with $\theta_0 = 2\arcsin(1/\gamma)=\theta(-1/\gamma)$. Furthermore
$d\theta/d\beta=0$ for $\beta=-1/\gamma$ producing a singularity in the
differential scattering cross section at $\theta_0$.

The differential cross section $d\sigma/d\theta$ gives the total
weight of impact parameters, which give scattering into the direction
$\theta$.
Equation~\ref{eq:scatt} has at most two solutions, which are easily
found to be
\begin{equation}
\beta_{\pm}(\theta)=-\gamma\sin^2 \theta/2\; \pm\;
       \cos\theta/2 \sqrt{1-\gamma^2\sin^2\theta/2}.
\end{equation}
Furthermore we have
\begin{equation}
\frac{d\beta_{\pm}}{d\theta}= -\frac{\gamma}{2} \sin\theta \mp
\sin\theta/2\frac{1+\gamma^2\cos\theta}{2\sqrt{1-\gamma^2\sin^2\theta/2}}.
\end{equation}
The differential cross section is for $0<\gamma<1$ given by
\begin{equation}
\frac{d\sigma}{d\theta}= \left\{
\begin{array}{lll}
r_0\displaystyle{\left|\frac{d\beta_-}{d\theta}\right|} &
\text{for} & \theta<0 \\
r_0\displaystyle{\left|\frac{d\beta_+}{d\theta}\right|} &
\text{for} & \theta>0,
\end{array}\right.
\end{equation}
and for $\gamma>1$, it is given by the expression
\begin{equation}
\frac{d\sigma}{d\theta}= \left\{
\begin{array}{lll}
r_0\displaystyle{\left|\frac{d\beta_-}{d\theta}\right|} +
r_0\displaystyle{\left|\frac{d\beta_+}{d\theta}\right|} &
\text{for} & 0<\theta<\theta_0 \\
0 & \text{otherwise}. &
\end{array}\right.
\end{equation}
Examples of cross-sections and trajectories are shown in Fig.~\ref{fig:ccs}.
The integrated cross section $\sigma_{\text{tot}}$
is equal to the total weight of impact parameters corresponding to particles
which hit the vortex. It is equal to the diameter of the vortex
$\sigma_{\text{tot}}=2r_0$, as is always the case in classical scattering.
If the classical cross section is substituted into the collision integral of
the Boltzmann transport equation, it is straightforward as shown in
App.~\ref{app:boltz}, to obtain expressions for the transverse and longitudinal
resistivities induced by a random distribution of vortices, in the limit where
the density of vortices goes to zero, thereby eliminating the effects of
multiple scattering events. Let $B$ denote the average flux density applied to
the sample, and $\rho^0_H = B/n_e e$ the Hall resistivity of a 2DEG in a
homogeneous magnetic field $B$. We then have
\begin{equation}
\frac{\rho_{xy}}{\rho^0_H} = \zeta^{\text{clas}}_{xy},\;\;\;\;\;\;\;
\frac{\rho_{xx} - \rho_{\text{imp}}}{\rho^0_H} = \zeta^{\text{clas}}_{xx},
\end{equation}
with
\begin{eqnarray}
\zeta^{\text{clas}}_{xy} &=& \frac{2\gamma}{r_0}\int^{\pi}_{-\pi}
 \frac{d\theta}{2\pi}
\sin\theta\frac{d\sigma}{d\theta}, \\
\zeta^{\text{clas}}_{xx} &=& \frac{2\gamma}{r_0}\int^{\pi}_{-\pi}
 \frac{d\theta}{2\pi}
(1-\cos\theta ) \frac{d\sigma}{d\theta}.
\end{eqnarray}
The $\zeta^{\text{clas}}$ parameters are a function of the sole parameter
$\gamma= l_c/r_0$. This is indeed what one would expect, since the transport
coefficients calculated within a classical framework can not depend on the
flux quantum $\phi_0$.
The functional behavior of the $\zeta$-quantities, are shown in
Fig.~\ref{fig:zclas}.
We will postpone the discussion of the experimental
consequences of these curves, until we have calculated the same quantities
quantum mechanically.

\subsection{Quantum Dynamics}\label{sec:quantum}

In this section we shall consider the electron scattering on a magnetic
flux vortex within the framework of quantum theory. We will
calculate the electron wave function, from which the longitudinal and
transverse conductivities can be found from the theory of
Sec.~\ref{sec:moment}. The quantum nature of the electron
radically alters the picture of the scattering
process, when the wavelength of the electron is comparable to, or longer
than the diameter of the vortex. In the limit of very small electron
wavelength, the scattering can essentially by described by the laws of
geometrical optics, and thereby classical mechanics.

\subsubsection{The Scattering Wave Function}
\label{sec:scatwave}

We will again take an idealized cylindrical vortex, with constant magnetic
field inside, and zero field outside, Eq.~\ref{eq:bfield}. This vortex is
completely symmetric under any rotation about the center axis. This symmetry
can also be made a symmetry of the Hamiltonian, by chosing the right gauge
when writing down the vector potential. In cylindrical coordinates
$\bbox{A}=\bbox{e}_r A_r + \bbox{e}_{\theta} A_{\theta}$, we have
\begin{equation}
B(\bbox{r})=\partial_r A_{\theta} -
  \frac{1}{r}\partial_{\theta}A_r +\frac{A_{\theta}}{r}.
\end{equation}
When $B$ is invariant under rotation, this equation has the simple solution
\begin{equation}
A_r=0,\;\;\;\;
A_{\theta}(r)=\frac{1}{r}\int_0^r dr'\, r'B(r'),
\end{equation}
which in our case give the vector potential
$\bbox{A}=\bbox{e}_{\theta} A_{\theta}$
with
\begin{equation}
A_{\theta}(r)=\left\{\begin{array}{ccc}
\displaystyle{\frac{\phi r}{2\pi r_0^2}}    &   \text{for} & r<r_0 \\
\displaystyle{\frac{\phi}{2\pi r}}      &  \text{for} & r>r_0.
\end{array}\right.
\end{equation}
The Hamiltonian is given by the expression
\begin{eqnarray}
H & = & \frac{1}{2m}\left(\bbox{p} - e\bbox{A}\right)^2  \nonumber \\
& = & -\frac{\hbar^2}{2m}\left\{\partial^2_r + \frac{1}{r}\partial_r +
\left(\frac{1}{r}\partial_{\theta} -
\frac{ie}{\hbar}A_{\theta}\right)^2\right\}.
\end{eqnarray}
Here we have taken the charge of the particle to be $e$. The Hamiltonian is
rotationally invariant, and therefore commutes with the angular momentum
about the symmetry axis, $L_z$. Consequently $L_z$ and $H$ have common
eigenstates. The canonical momentum operator of a particle in a magnetic
field is
$\bbox{p}=m\bbox{v}+q\bbox{A}=\frac{\hbar}{i}\bbox{\nabla}$, and the operator
for
the angular momentum about the $z$-axis is $L_z=[ \bbox{r\times p}]_z=
\frac{\hbar}{i}\partial_{\theta}$. The eigenstates of $L_z$ are $e^{il\theta}$,
where
$l$ must be an integer in order not to have cuts in the wave function.  We can
now
separate the variables of the common eigenstates of $L_z$ and $H$, and write
\begin{equation}
\phi_{kl}(r,\theta)=R_{kl}(r)e^{il\theta},
\end{equation}
where $k$ is an energy label $E=\hbar^2 k^2/2m$.
Let us introduce the flux quantum $\phi_0 =h/e$ and the dimensionless
fraction $\alpha=\phi/\phi_0$. The differential equations
for the radial part of
the wave function, takes a particularly simple form if we write them down in
dimensionless variables $\xi=r/r_0$ and $\kappa=k r_0$. The
variable $\kappa$ measures the size of the vortex compared to the
electron wavelength. With these definitions, the equation for the radial
part of the wave function for $\xi<1$ is
\begin{equation}
R'' + \frac{1}{\xi}R' +\left(\kappa^2 -
   (\frac{l}{\xi} - \alpha\xi)^2\right) R =0.
\end{equation}
And for $\xi>1$ we have
\begin{equation}
R'' +\frac{1}{\xi}R' +\left(\kappa^2 - \frac{(l-\alpha)^2}{\xi^2}\right) R=0.
\end{equation}
Inside the vortex an analytical solution to the radial equation can be found
by following the procedure of L. Page~\cite{Page30}, see also the review of
Olariu and Popescu~\cite{Rumaensk85}.
The solution to the radial equation inside the vortex, which is regular as
$\xi\rightarrow 0$, is
\begin{equation}
\phi_{k l}(\xi,\theta)=
C \xi^{|l|}e^{-\frac{1}{2}\alpha\xi^2}
M\left[\frac{1}{2}(|l|+1 -l - \frac{(k r_0)^2}{2\alpha}),|l|+1,
\alpha\xi^2 \right]
e^{il\theta}.
\end{equation}
Here $M$ is a confluent hypergeometric function (solution to Kummer's
equation), in the notation of Abramowits and Stegun~\cite{AS},
and $C$ is a normalization constant which we will not need to evaluate.
At the Landau quantization energies the confluent hypergeometric function $M$
reduces to an associated Laguerre polynomial.

Outside the vortex the
radial equation is just the differential equation for Bessel
functions of the first kind. We have for $\xi>1$
\begin{equation}
\phi_{k l}(\xi,\theta)=
\left(a_l J_{l-\alpha}(k r) + b_l Y_{l-\alpha}(k r)\right)e^{il\theta}.
\label{waveJY}
\end{equation}
The two constants $a_l$, $b_l$ are found from the requirement, that the
wave function has to be continuously differentiable at the boundary of the
vortex. Let us write
\begin{equation}
a_l=c_l \cos\delta_l\;\;\;\;\;\;\;
b_l=-c_l \sin\delta_l,
\end{equation}
then we have the following asymptotic expression for the total wave function
$\psi(r,\theta)$ in the region $kr\gg 1$
\begin{equation}
\psi(r,\theta)=\sqrt{\frac{2}{\pi k r}}\sum_l c_l\cos(x_l + \delta_l)
e^{il\theta}.
\label{eq:wtot}
\end{equation}
Here we have introduced the abbreviation $x_l=k r - (\pi/2)(l-\alpha)-\pi/4$.
The scattering phase shift $\delta_l$, is found from the
condition that the wave function has to be continuously differentially at the
boundary of the vortex. The constant $c_l$ can be expressed in terms of the
phase shift, via the scattering condition, i.e.\ that all the incoming
particle current must be described by the wave function $\psi_i$ which far
away from the vortex describes a uniform current directed towards the vortex.
In the geometry with which we are concerned, the particles come in along the
positive $x$-axis, towards the vortex placed at the origin, as shown in
Fig.~\ref{fig:qgeo}. The state
describing a uniform current in the direction of the negative $x$-axis is
\begin{equation}
\psi_i=e^{-ikr\cos\theta + i\alpha\theta}.
\end{equation}
This expression is observed to have a cut whenever $\alpha$ is not integer,
however this cut will not be present in the
total wave function $\psi$. There will be a counter cut in the
outgoing wave function $\psi_s=\psi - \psi_i$, to render the total wave
function completely smooth. This cut only arises because we insists in
splitting the total wave function into incoming and outgoing terms. We take
the angle $\theta$ to have values in the interval $-\pi<\theta<\pi$, which
places the cuts in $\psi_{i,s}$ along the negative real axis. With this
choice the asymptotic expansion of $\psi_i$, for $k r\gg 1$,
in terms of partial waves, takes the form
\begin{equation}
\psi_i=\frac{1}{\sqrt{2\pi k r}}\sum_l \left[
e^{ikr-i\pi/4}\cos\pi(l-\alpha) + e^{-ikr +i\pi/4}\right].
\end{equation}
The scattering condition is now implemented by demanding that the coefficient
of $e^{-ikr}$, in respectively the expansion of $\psi$ and $\psi_i$ must be
identical for each $l$, resulting in the identity
\begin{equation}
c_l=e^{i\delta_l - i\pi(l-\alpha)/2}.
\end{equation}

\subsubsection{Phase Shifts for Scattering on Vortex with Finite
Radius}\label{subsec:phase}

Let us now calculate the phase shifts for a charged particle scattering on a
cylindrical flux vortex with finite radius.
The phase shifts are derived from
the condition that the logarithmic derivative of the radial wave function must
be continuous at the boundary of the vortex
\begin{equation}
\frac{1}{R_l^{<}}\left.\frac{d R_l^<}{d\xi}\right|_{\xi=1}=
\frac{1}{R_l^>}\left.\frac{d R_l^{>}}{d\xi}\right|_{\xi=1}.
\end{equation}
Inside the vortex we have with the abbreviations
$h_l \equiv (|l|+1-l-(kr_0)^2/2\alpha)/2$ and $g_l \equiv |l|+1$
\begin{equation}
E_l  \equiv  \frac{1}{R_l^{<}}\left.\frac{d R_l^<}{d\xi}\right|_{\xi=1}
= |l| - \alpha + 2\alpha\frac{h_l}{g_l}\,
\frac{M[h_l+1,g_l+1,\alpha]}{M[h_l,g_l,\alpha]}.
\end{equation}
Outside the vortex the logarithmic derivative reads
\begin{equation}
\frac{1}{R_l^>}\left.\frac{d R_l^{>}}{d\xi}\right|_{\xi=1}=
\frac{j_l - y_l \tan\delta_l}
{J_{l-\alpha}(\kappa) - Y_{l-\alpha}(\kappa)\tan\delta_l},
\end{equation}
where we have introduced the abbreviations
$z_l=(kr_0/2)(Z_{l-\alpha-1}(kr_0) - Z_{l-\alpha+1}(kr_0))$ with
$(z,Z)=(j,J)$ or $(y,Y)$. It is now  simple to solve for $\delta_l$
\begin{equation}\label{eq:tangens}
\tan\delta_l =\frac{j_l - E_l J_{l-\alpha}(kr_0)}
{y_l - E_l Y_{l-\alpha}(kr_0)}.
\end{equation}
The $\tan\delta_l$'s are the bricks from which cross sections and transport
coefficients can be build.
The presented curves have all been calculated with phase
shifts found by this expression with the help of {\em Mathematica}, which have
implementations of all the involved special functions.

\subsubsection{Momentum Flow and Force Balance}
\label{sec:moment}

In order to express the transverse and longitudinal
resistivities of the 2DEG in terms of matrix elements of the force,
we make a simple force balance argument.
For a more comprehensive discussion of the
force balance, we refer to E. B. Hansen~\cite{Brun}.

We consider a 2DEG carrying a current of density $j_x$, along the
$x$-direction, i.e.\ $j_y=0$.
The distribution function
of the electrons is then
\begin{equation}
n(\bbox{k})=n^0(\bbox{k} - \delta\bbox{k}),
\end{equation}
with $n^0(\bbox{k})=\Theta(\epsilon_F - \epsilon_{\bbox{k}})$, and
$\delta\bbox{k}$ is related to the current density $j_x=(ev^2_F/2\hbar)
g(\epsilon_F)\delta k$, where $g$ is the density of states. The forces
acting on an electron in the state $\bbox{k}$ can be separated into two
classes. First, there are the forces due to the electric fields $E_x$, $E_y$
present in the sample. Secondly, there are the forces due to scattering
processes, which due to the rotational symmetry of the scatterers -- vortices
and impurities -- can be decomposed into a force $F_L$ parallel to $\bbox{k}$,
and a force $F_T$ at right angles to $\bbox{k}$. If we let $\theta_{\bbox{k}}$
denote the angle between $\bbox{k}$ and the $x$-axis, we can write the
scattering force on an electron in the state $\bbox{k}$, projected along the
coordinate axes as
\begin{eqnarray}
F_x(\bbox{k}) & = & F_L(k)\cos\theta_{\bbox{k}} - F_T(k)\sin\theta_{\bbox{k}},
\\
F_y(\bbox{k}) & = & F_L(k)\sin\theta_{\bbox{k}} + F_T(k)\cos\theta_{\bbox{k}}.
\end{eqnarray}
When we have a steady state transport situation, the total force on the 2DEG
must vanish. With $\mu =x,y$, and $n_e$ denoting the density of the 2DEG,
we therefore have the equation
\begin{equation}
e n_e E_{\mu}    =  \sum_{\bbox{k}}F_{\mu}(\bbox{k})n(\bbox{k}).
\end{equation}
And thereby
\begin{equation}
e n_e E_{\mu} =  \frac{v_F}{\hbar}g(\epsilon_F)\delta k
\int\frac{d\theta_{\bbox{k}}}{2\pi}
F_{\mu}(k_F,\theta_{\bbox{k}})\cos\theta_{\bbox{k}}.
\end{equation}
Along the $x$- and $y$-directins respectively, this leads to
\begin{eqnarray}
\frac{E_x}{j_x} & = & \frac{F_L(k_F)}{n e^2 v_F} \\
\frac{E_y}{j_x} & = & \frac{F_T(k_F)}{n e^2 v_F}.
\end{eqnarray}
The ordinary impurities etc.\ do not give rise to a net transverse force on
the electrons. Consequently, only the vortices contribute to the transverse
force $F_T(k_F)$ on an electron, on the Fermi surface. This force will, in
the limit of low vortex density, be proportional to the total number of
vortices. In this limit we can therefore calculate the force as the total
number of vortices $N_{\alpha}$, times the force
from a single vortex
\begin{equation}
F_T(k_F) = N_{\alpha} \left\langle\Psi |\hat{F}_{y}| \Psi\right\rangle
\end{equation}
Here $N_{\alpha}=An_{\alpha}=AB/\alpha\phi_0$, with $B$ equal to the average
flux density applied to the sample.
Further more,
$\hat{\bbox{F}}= (e/2)[\bbox{v}\times\bbox{B}+ \bbox{B}\times\bbox{v}]$
is the operator corresponding to the Lorentz force inside the vortex,
and $\Psi$ is the normalized wave function of the electron, i.e.\
$\left\langle\Psi | \Psi\right\rangle=1$. The electron wave function $\Psi$
is an eigenstate of the one electron -- one vortex system. It
describes an electron which initially travels  along the $x$-axis, or in other
words, the expectation value of the velocity is
$\left\langle\Psi |\hat{v}_{\mu}|\Psi\right\rangle=\delta_{\mu x}v_F$, and
it is labeled by the k-vector $\bbox{k}=(k_F,0)$.
Note, that $\Psi$ is identical to the wave function
$\psi$ of Sec.~\ref{sec:scatwave} apart from the normalization
$\left\langle\psi | \psi\right\rangle=A$.
The transverse resistivity of the 2DEG can now be expressed as
\begin{equation}
\rho_{xy}=\frac{E_y}{j_x}=\rho^0_{H}\zeta_{xy},
\end{equation}
where $\rho^0_{H}=B/n_ee$ is the Hall resistivity of electrons in a homogeneous
magnetic field of magnetic flux density $B$,
and
\begin{equation}
\zeta_{xy}=\frac{\left\langle\psi |\hat{F}_{y}| \psi\right\rangle}
{2\pi\alpha\hbar v_F}.
\end{equation}

Similarly, we have for the longitudinal resistivity
\begin{equation}
\rho_{xx} - \rho_{\text{imp}}=\rho^0_{H}\zeta_{xx}=
\frac{\hbar}{e^2}\frac{n_{\alpha}}{n_e}2\pi\alpha\zeta_{xx}.
\end{equation}
where we have singled out the contribution from the vortices, and
\begin{equation}
\zeta_{xx}=\frac{\left\langle\psi |\hat{F}_{x}| \psi\right\rangle}
{2\pi\alpha\hbar v_F}.
\end{equation}

The matrix elements $\left\langle\psi |\hat{F}_{\mu}| \psi\right\rangle$ can be
calculated by numerically performing the integral, which is limited to the
range of the magnetic field, i.e.\ to the core of the vortex. We have used
this procedure, which is rather laborious for the computer, to check the
results obtained by the following much faster procedure.

In a continuum the
flow of kinematical momentum is described by a continuity equation, which at
the classical level takes the form
\begin{equation}
\partial_t k_{\mu} + \partial_{\nu}\pi_{\nu \mu} = F_{\mu}.
\label{eq:i}
\end{equation}
Here $k_{\mu}=mv_{\mu}$ is the kinematical momentum (which in the presence of
a magnetic field differ from the generalized momentum
$p_{\mu}=mv_{\mu}+eA_{\mu}$) and $\pi_{\nu \mu}=mv_{\nu}v_{\mu}$ is
the tensor describing the flow of kinematical momentum.
The quantum mechanical equivalent of Eq.~\ref{eq:i} is
\begin{equation}
\partial_t k_{\mu} + \partial_{\nu}\pi_{\nu \mu} =
\Psi^* \hat{F}_{\mu} \Psi,
\label{eq:ii}
\end{equation}
where now
\begin{eqnarray}
k_{\mu} &=& m\Psi^* v_{\mu} \Psi, \\
\pi_{\nu \mu} &=& \frac{m}{2}\left[
\Psi^* (v_{\nu}v_{\mu}\Psi) + (v_{\mu}\Psi)(v_{\nu}\Psi)^*\right],
\end{eqnarray}
and $\hat{\bbox{F}}$ is the
operator corresponding to the Lorents force.
The interpretation of Eq.~\ref{eq:ii} as a generalization of Eq.~\ref{eq:i}
applies strictly speaking only to the real part. For a derivation of
Eq.~\ref{eq:ii} we refer to the review of
Olariu and Popescu~\cite{Rumaensk85}.
The scattering situation we are considering is in a steady state and
therefore $\partial_t k_{\mu}=0$. We have by Gauss law
\begin{equation}
\left\langle\psi |\hat{F}_{\mu}|\psi\right\rangle =
\int d^2 r\, \partial_{\nu}\pi_{\nu \mu}(\bbox{r})
=\int_{-\pi}^{\pi}r d\theta\, \pi_{r \mu}(\bbox{r}),
\label{eq:oint}
\end{equation}
where the index $r$ refers to the radial component, and the last integral is
along any circle of radius $r>r_0$,
(the Lorentz force is only non-vanishing inside the vortex).
Let us first consider the $y$-component of
the force. As $\pi_{\nu \mu}$ is a tensor, we can write
$\pi_{r y}=\pi_{rr}\sin\theta + \pi_{r\theta}\cos\theta$. Furthermore the
velocity operators are (outside the vortex core)
\begin{eqnarray}
v_r &=& \frac{\hbar}{m}\frac{\partial_r}{i} \\
v_{\theta} &=& \frac{\hbar}{m r}(\frac{\partial_{\theta}}{i} - \alpha),
\end{eqnarray}
and due to the extra factor of $r$ in $v_{\theta}$,
only $\pi_{rr}$ will contribute to
the integral, Eq.~\ref{eq:oint}, in the limit $r\rightarrow\infty$. In the
asymptotic region we have
\begin{equation}
\pi_{rr} = \frac{\hbar^2 k}{\pi m r}\sum_m \sum_l
e^{i(\delta_{l+m} - \delta_l) - i\frac{\pi}{2}m}
\cos[\delta_{l+m} - \delta_l - \frac{\pi}{2}m]e^{im\theta},
\label{eq:pirr}
\end{equation}
and
\begin{equation}
\left\langle\psi |\hat{F}_y | \psi\right\rangle =
\int_{-\pi}^{\pi}r d\theta\,\pi_{rr}\sin\theta,
\end{equation}
resulting after some algebra in
\begin{equation}
\zeta_{xy} = \frac{1}{2\pi\alpha}\sum_l\sin[2(\delta_{l+1} - \delta_l)].
\label{eq:zxy}
\end{equation}
Similarly, we have in the $x$-direction $\pi_{r x}=\pi_{rr}\cos\theta -
\pi_{r\theta}\sin\theta$, and
\begin{equation}
\left\langle\psi |\hat{F}_x| \psi\right\rangle =
\int_{-\pi}^{\pi} r d\theta\,\pi_{rr} \cos\theta,
\end{equation}
resulting in the expression
\begin{equation}
\zeta_{xx}=\frac{1}{\pi\alpha}\sum_l \sin^2 [\delta_{l+1} - \delta_l].
\label{eq:zxx}
\end{equation}
In order to calculate the $\zeta$-quantities it is convenient to express them
in terms of $t_l\equiv\tan\delta_l$
\begin{eqnarray}
\zeta_{xx} &=& \frac{1}{\pi\alpha}\sum_l
\frac{(t_{l+1}-t_l)^2}{(1+t_l^2)(1+t_{l+1}^2)}, \\
\zeta_{xy} &=& \frac{1}{\pi\alpha}\sum_l
\frac{(t_{l+1}-t_l)(1+t_lt_{l+1})}{(1+t_l^2)(1+t_{l+1}^2)}.
\end{eqnarray}
The $\zeta_{xy}$ expression can be divided into two terms of different origin.
Define $\Delta\zeta_{xy}$ and $\Delta_{xy}(\alpha)$ to be respectively
\begin{eqnarray}
\Delta\zeta_{xy} &\equiv& \frac{2}{\pi\alpha}\sum_l
\frac{(t_{l+1}-t_l)t_lt_{l+1}}{(1+t_l^2)(1+t_{l+1}^2)}, \\
\Delta_{xy}(\alpha) &\equiv& \zeta_{xy} - \Delta\zeta_{xy} =
\frac{(t_{l+1}-t_l)(1-t_lt_{l+1})}{(1+t_l^2)(1+t_{l+1}^2)}.
\end{eqnarray}
Then we have
\begin{eqnarray}
\Delta_{xy}(\alpha) &=& \frac{1}{\pi\alpha}\sum_l
\left[\frac{t_{l+1}}{1+t_{l+1}^2} - \frac{t_l}{1+t_l^2}\right] \nonumber \\
&=& \frac{1}{\pi\alpha}\left[\sin\delta_{\infty}\cos\delta_{\infty} -
\sin\delta_{-\infty}\cos\delta_{-\infty}\right] \nonumber \\
&=& \frac{\sin 2\pi\alpha}{2\pi\alpha}.
\label{dxy}
\end{eqnarray}
Here we have used the fact that the sum is telescopic, and the following
properties of the phase shifts. For any given finite energy of the electron,
the $l$'th partial wave, which is an eigenstate of $L_z$ with eigenvalue
$\hbar l$, will have an insignificantly small propability density inside a
radius $b_l$, determined by $\hbar |l|\sim b_l \hbar k$, i.e.\ $b_l\sim
|l|/k$. As $|l|$ grows to infinity the $l$'th partial wave will not be able
to detect any difference between a vortex of radius $r_0\ll b_l$, and an
Aharonov-Bohm string, and consequently $\delta_l \rightarrow \delta_l^{AB}$
for $l \rightarrow\pm\infty$. The Aharonov-Bohm phase shifts are
\begin{equation}
\delta_l^{AB} = \left\{ \begin{array}{ccc}
0 & \text{for} & l>\alpha \\
\pi(l - \alpha) & \text{for} & l<\alpha, \end{array}\right.
\end{equation}
which can be seen by comparing  the wave functions given respectively in
Eq.~\ref{waveJY} and Eq.~\ref{wAB}. By inserting the Aharonov-Bohm phase
shift into Eq.~\ref{dxy}, the result follows
\begin{equation}
\zeta_{xy} = \frac{\sin 2\pi\alpha}{2\pi\alpha} +
\frac{(t_{l+1}-t_l)t_lt_{l+1}}{(1+t_l^2)(1+t_{l+1}^2)}.
\end{equation}
By inserting $\delta_l^{AB}$ into $\Delta\zeta_{xy}$, it is seen that
$\Delta\zeta_{xy}$ vanishes in this limit. In this form $\zeta_{xy}$ is
therefore separated into a term due to the Aharonov-Bohm effect, and a
correction term due to the effects of the finite radius of the vortex.
Moreover the sum over $l$ in the correction term, converges rapidly for
increasing $|l|$. The $\Delta_{xy}(\alpha)$ term show, that even though the
magnetic field is only non-vanishing at a single point in the Aharonov-Bohm
limit, the electron still experiences a transverse force. This force was first
described by S. V. Iordanskii~\cite{Iordanskii}.
Let us remark that the sum appearing in $\zeta_{xx}$ is
already convergent. The residual value of $\zeta_{xx}$ in the
Aharonov-Bohm limit is
\begin{equation}
\zeta_{xx}^{AB}=\frac{1}{\pi\alpha}\sin^2\pi\alpha.
\end{equation}

Curves showing $\zeta_{xy}$ as a function of $k r_0$ for different values of
the flux $\alpha$, have been plotted in Fig.~\ref{fig:zxy}.
For $k r_0\gg 1$ all the curves go to $\zeta_{xy}=1$, which is the classical
value as is seen in Fig.~\ref{fig:zclas}. In the Aharonov-Bohm limit
$k r_0\ll 1$, the curves go to a residual value depending on the flux
$\zeta_{xy}\rightarrow\zeta_{xy}^{AB}=\sin 2\pi\alpha/2\pi\alpha$.
The $\zeta_{xy}$-curve for $\alpha=1/2$ we can compare with the experimental
Hall factor measured by Geim et al.\ \cite{Geim92}.
The overall qualitative behavior is in good agreement.
To make a quantitative test we have fitted our curve to the experimental curve
by tuning the radius of the vortex $r_0$.
The best fit is obtained for a vortex radius $r_0=50$nm, and
this is significantly smaller than the exponential length
estimated by Geim to be 100nm.

Curves showing $2\pi\alpha\zeta_{xx}$ as a function of $k r_0$ for different
values of the flux $\alpha$, is shown in Fig.~\ref{fig:zxx}.
The magneto-resistivity, given by
\begin{equation}
\frac{d\rho_{xx}}{dB}=\frac{1}{ne}\zeta_{xx},
\label{eq:neto}
\end{equation}
is constant, corresponding to the fact that the resistivity is proportional
to the density of vortices, in the low $B$-field limit. The magnitude of the
magneto-resistivity do not show perfect agreement with the values measured by
Geim et al.~\cite{Geim93}. For the 2DEG with the lowest density that has been
measured by Geim, $n_e = 3.65\cdot 10^{10}\text{cm}^{-2}$, Geim reports
$d\rho_{xx}^{\text{meas}}/dB = 0.06\Omega/G$. In this case Eq.~\ref{eq:neto}
predicts $d\rho_{xx}/dB = 0.55\Omega/G$, for $r_0=50$nm.

In order to investigate the sensitivity of
the $\zeta$'s to the radial distribution of magnetic field in the vortex, we
have calculated them with an exponential distribution
\begin{equation}
B_{\text{exp}}(r)=\frac{\alpha\phi_0}{2\pi r_0^2} e^{-r/r_0}.
\end{equation}
The calculation is performed by solving the radial Shr\"{o}dinger equation
numerically, and imposing an artificial boundary on the vortex, outside which
the amount of flux is negligible. The phase shifts are then found from
the condition that the wave function much be smooth at the boundary. From the
phase shifts the $\zeta$-quantities can be calculated as for the step function
vortex. The calculation shows almost no difference between the $\zeta$-values,
corresponding to the two different choices, indicating that the
$\zeta$-quantities are not very sensitive to the shape of the vortices.

\subsection{Multi Flux Quantum Vortex and Resonance \- Scattering}
\label{sec:reso}

When the total amount of magnetic flux inside the vortex is increased, the
$\zeta$-quantities acquire more structure.
To illustrate this we have plotted $\zeta_{xy}$ and $2\pi\alpha\zeta_{xx}$ in
Fig.~\ref{fig:reso} for a flux vortex carrying a total of
$\alpha=$10 flux quanta.
The structure seen in the plots is an effect of the resonant scattering
which takes place when the energy of the incoming particle is
close to one of the Landau quantization energies corresponding to the
magnetic field strength inside the vortex. The magnetic field in the vortex
vanishes outside a finite range -- the radius of the vortex -- and there
are therefore no real Landau levels in the sense of stationary eigenstates,
but only metastable states.
The resonance condition $(p+1/2)\hbar\omega_c=\hbar^2 k^2/2m$ translates for
fixed $\alpha$ into
\begin{equation}
k_p r_0=2\sqrt{|\alpha|(p+\frac{1}{2})},\hspace{0.5cm}p=0,1,2,\dots
\end{equation}
These values are in excellent agreement with the resonances seen in
Fig.~\ref{fig:reso}, where the first eight resonances corresponding to
$p=0,\dots,7$, are clearly distinguished.
At the resonance energies the typical time the particle spends
in the scattering region, i.e.\ inside the vortex, is much longer than it
is away from resonance. The time the particle spends inside the vortex at
a resonance, is the lifetime of the corresponding metastable
state. The inverse lifetime is proportional to the width of the resonance,
that is strictly speaking the width of the peak in the partial wave cross
section $\sigma_l$, corresponding to the $l$ quantum number of the
metastable Landau state.

It is easy to interpret the peaks in the $\zeta_{xx}$-curve in
Fig.~\ref{fig:reso}, appearing at the resonance energies. Because when the
electron spends longer time in the scattering region, it loses knowledge of
where it came from, resulting in an enhanced probability of being scattered
in the backwards direction. The $\zeta_{xy}$-curve is a measure of asymmetric
scattering, and we can therefore interpret the dips seen in
Fig.~\ref{fig:reso} along the same line of reasoning.
The electron spends longer time in the scattering region,
thereby losing knowledge of what is left and what is right.
Every time a new ``channel'' is opened the asymmetry of the scattering is
suppressed, thus resulting in a sawtooth like curve.

\section{Hall Effect in a Regular Array of Flux Vortices}\label{sec:array}

\subsection{Introduction}
\label{sec:intro}

Recently measurements were made by Geim et al.\ \cite{Geim92,Geim93} of the
Hall resisti\-vi\-ty of low density 2DEG's in a random distribution of flux
vortices, at very low magnetic field strengths. A profound suppression
of the Hall resistivity was found, for 2DEG's with Fermi wavelengths
of the same order of magnitude as the diameter of the flux vortices.
This indicates that we are dealing with a phenomenon of quantum nature.
These measurements were made by placing a thin lead film on top of a
GaAs/GaAlAs heterostructure.
When a perpendicular magnetic field is applied,
the magnetic field penetrates the superconducting lead film and also
the heterostructure in the form of flux vortices
each carrying half a flux quantum $\phi_0 /2$ of magnetic flux. Due to the
strong flux vortex pinning in the films Geim have used, the vortices
were positioned in a random configuration.

In this section we will consider the hypothetical experiment where one
instead of a ``dirty'' film, places a perfectly homogeneous type II
superconducting film, on top of the 2DEG.
The film do not have to be made of a material
which is type II superconducting in bulk form. A film of a type I
superconducting material will also display a mixed state if the thickness of
the film is below the critical thickness $d_c$. Experimentally perfect
Abrikosov flux vortex lattices have been observed in thin films of lead
with thickness $d<d_c\cong 0.1\mu$m, \cite{Hyb}.

If one succeeds to make such a sandwich construction, one has an ideal
system for investigating how a 2 dimensional electron gas behaves in a
periodic magnetic field. When the magnetic field exceeds $H_{c1}$, which can be
extremely low, the superconducting film will enter the mixed phase, and
form an Abrikosov lattice of flux vortices. The
Abrikosov lattice in the superconductor will give rise to a periodic magnetic
field at the 2DEG, and moreover as the strength of the applied magnetic field
is varied the only difference at the 2DEG, is that the lattice constant of
the periodic magnetic field varies. The Abrikosov lattice is most often a
triangular lattice with hexagonal symmetry, although other lattices have
been observed (e.g.\ square) in special cases where the atomic lattice
structure impose a symmetry on the flux lattice, \cite{Hyb}.
For simplicity, the model calculations we have made are for a system with a
square lattice
of flux vortices, but we do not expect this to influence the overall
features of the results. From the point of view of the 2DEG it is
important that the flux vortices carry half a flux quantum
$h/2e = \phi_0/2$ due to the $2e$-charge of the Cooper pairs
in the superconductor. The magnetic field from a single flux vortex fall of
exponentially with the distance from the center of the vortex. This
exponential decay is characterized by a length $\lambda_s$, which essentially
is the London length of the superconductor, proportional to one over the
square root of the density of Cooper pairs. The length $\lambda_s$ can be
varied by changing the temperature, or the material of the superconductor.

The other characteristic lengths of the system are the Fermi wavelength
$\lambda_F =\sqrt{2\pi/n_e}$, where $n_e$ is the density of the 2DEG,
the lattice constant $a$ of the periodic magnetic field, and the mean free
path $l_f=v_F \tau$. The mean free path we assume to be very large compared to
$a$ and $\lambda_s$. The magnetic field is only varying appreciably when $a$
is larger than $\lambda_s$. This means that the magnetic flux density of the
applied field should be appreciably less than $\phi_0/(\pi\lambda_s^2)$.
 In the limit where
$\lambda_s \ll a,\lambda_F$, the vortices
can be considered magnetic strings, and the electrons experiences a periodic
array of Aharonov-Bohm scatterers.
In this case the value of the flux through each
vortex is crucial. If for instance the flux had been one flux quantum
$\phi_0$, the electrons would not have been able to feel the
vortices at all. But in the real world the vortices from the superconductor
carry $\phi_0/2$  of flux, and therefore this limit is
nontrivial. The electrons has for instance a band structure quite different
from that of free electrons. In the mathematical limit of infinitely thin
vortices each carrying {\em half} a flux quantum, there cannot be any
Hall effect. This is most easily seen by subtracting one flux quantum
from each vortex to obtain a flux equal to minus a half flux quantum
through each vortex. As we have discussed earlier the introduction of the
Dirac strings can not change any physics, and the procedure therefore shows
that the system is equivalent to it's time reversed counterpart, thereby
eliminating the possibility of a Hall effect.

In this study we have ignored electron-electron interaction and
the electron spin throughout, in order to keep
the model simple. From the point of view of the phenomena we are going to
describe, the effect of the electron spin will be to add various small
corrections.

\subsection{Electrons in a periodic magnetic field}
\label{sec:elec}

\subsubsection{Magnetic translations}

It is a general result for a charged particle in a spatially periodic
magnetic field $B(x,y)$, that the eigenstates of the system can be labeled by
Bloch vectors taken from a Brillouin zone, if and only if the flux through
the unit cell of the magnetic field is a rational number $p/q$ times the
flux quantum.
The standard argument for this fact is made by introducing magnetic
translation operators.
To introduce magnetic translation operators in an inhomogeneous
magnetic field we first make the following observation.
The periodicity of the field can be stated $B(\bbox{r}+\bbox{R})=B(\bbox{r})$,
for $\bbox{R}$ belonging to a Bravais lattice. But this implies that the
difference between the vector potentials $\bbox{A}(\bbox{r}+\bbox{R})$ and
$\bbox{A}(\bbox{r})$ must be a gauge transformation
\begin{equation}
\bbox{\nabla}\times\left\{\bbox{A}(\bbox{r}+\bbox{R}) -
\bbox{A}(\bbox{r})\right\}=
B(\bbox{r}+\bbox{R}) - B(\bbox{r}) =0.
\end{equation}
We introduce the gauge potential $\chi_{\bbox{R}}$ and write
\begin{equation}
\bbox{A}(\bbox{r}+\bbox{R}) = \bbox{A}(\bbox{r}) +
\bbox{\nabla}\chi_{\bbox{R}}(\bbox{r}).
\end{equation}
The function $\chi_{\bbox{R}}$ is only defined modulo an arbitrary
additive constant
which have no physical effect.
The Hamiltonian of the electrons is
\begin{equation}\label{eq:ham}
H = \frac{1}{2m}\left(\bbox{p} - e\bbox{A}(\bbox{r})\right)^{2}.
\end{equation}
The ordinary translation operators
$T_{\bbox{R}}=\exp[\frac{i}{\hbar}\bbox{R}\cdot\bbox{p}]$ do not commute
with the Hamiltonian, because they shift the
argument of the vector potential from $\bbox{r}$ to $\bbox{r}+\bbox{R}$, but as
we just have seen this can be undone with a gauge transformation. We therefore
introduce the magnetic translation operators as the combined symmetry
operation of an ordinary translation and a gauge transformation
\begin{equation}
M_{\bbox{R}}=\exp[-i\frac{e}{\hbar}\chi_{\bbox{R}}(\bbox{r})]
      \exp[\frac{i}{\hbar}\bbox{R}\cdot\bbox{p}].
\end{equation}
The operator $M_{\bbox{R}}$ is unitary, as it is the product of two unitary
operators, and therefore has eigenvalues of the form $e^{i\lambda}$. Let us
denote the primitive vectors of the Bravais lattice $\bbox{a}$ and $\bbox{b}$.
We can find common eigenstates of $M_{\bbox{a}}$, $M_{\bbox{b}}$ and $H$,
if and only if they all commute with each other. The magnetic translations
each commute with the Hamiltonian by construction, and furthermore we have
\begin{equation}
M_{\bbox{a}}M_{\bbox{b}} =  \exp[2\pi i\frac{\phi}{\phi_0}]
M_{\bbox{b}}M_{\bbox{a}}.
\end{equation}
If the flux $\phi$ through
the unit cell is a rational number $p/q$ ($p$ and $q$ relatively prime) times
the flux quantum $\phi_0$, $M_{q\bbox{a}}$ and $M_{\bbox{b}}$ commute. In this
case the cell spanned by $q\bbox{a}$ and $\bbox{b}$ is called the magnetic
unit cell. Let us define $\bbox{c}=q\bbox{a}$. The possible eigenvalues of
$M_{\bbox{c}}$ are phases $e^{2\pi i k_1 }$, where we can restrict
$|k_1 |<\frac{1}{2}$, and equivalently for $M_{\bbox{b}}$. We can therefore
label the common eigenstates
$|\bbox{k},n\rangle$, where $\bbox{k}=k_1 \bbox{c}^* + k_2\bbox{b}^*$, and
$\bbox{c}^*,\bbox{b}^*$ are the primitive vectors of the reciprocal lattice.
The vector $\bbox{k}$ is restricted to the magnetic Brillouin zone.
An arbitrary magnetic translation of an eigenstate with a Bravais lattice
vector $\bbox{R}=n\bbox{c}+m\bbox{b}$ can now be written
$M_{\bbox{R}}|\bbox{k},n\rangle=
(M_{\bbox{c}})^n (M_{\bbox{b}})^m |\bbox{k},n\rangle
=\exp[i \bbox{k}\cdot\bbox
{R}]|\bbox{k},n\rangle$, showing that the eigenstate
is a Bloch state. In this case we can speak of
energy bands forming a band structure in the usual sense. When the flux
through the elementary unit
cell $(\bbox{a},\bbox{b})$ is an irrational number of flux quanta, the
situation
is different. The irrational number can be reached as the limit where $p$ and
$q$ get very large, and consequently the Brillouin zone get very small and
collapses in the limit.

\subsubsection{The Dirac vortex viewpoint}

In this section we will show, how it is possible to argue in a slightly
different way from the previous section, and hereby in a simpler way  obtain
the vector potential of a periodic magnetic field.
 Let us again assume a rectangular unit cell
$(a,b)$, $B(x+a,y)=B(x,y+b)=B(x,y)$ etc., to keep the notation simple.
The magnetic field enter the Hamiltonian only through the vector potential.
The question is therefore if one can choose a gauge such that the
vector potential will be translationally invariant relative to a unit cell
$(c,d)$ $\bbox{A}(x+c,y)=\bbox{A}(x,y+d)=\bbox{A}(x,y)$ etc. It is clear that
if
such a periodic $\bbox{A}$-field exists, then the total flux $\Phi_{cd}$
through the unit cell $(c,d)$ will be zero, as it is given by the line
integral of $\bbox{A}$ around the boundary of the unit cell $(c,d)$
\begin{equation}
\Phi_{cd}=\oint_{\partial (c,d)} \bbox{A}\cdot\bbox{dl},
\end{equation}
which is zero by the periodicity.
We remark that due to the relation $B=\bbox{\nabla}\times\bbox{A}$,
the cell $(c,d)$ will be bigger than or equal to $(a,b)$. If the flux
through the unit cell of the magnetic field is not zero, but equal to a
rational number times the flux quantum, $\Phi_{ab}=p/q\;\phi_0$ ($p$ and $q$
relatively prime), a trick can be applied to make the flux $\Phi_{cd}$
become zero.
It is a basic fact, apparently first observed by Dirac~\cite{Dirac},
that a particle with charge $e$ cannot feel an infinitely thin solenoid
carrying a flux equal to an integer multiple of the flux quantum
$\phi_{0}=h/e$.
Such a stringlike object carrying one flux quantum is sometimes called a
Dirac vortex.
To find the periodic vector potential take an enlarged unit cell
$(c,d)=(qa,b)$, so that $\Phi_{cd}=p\phi_0$ and
put by hand $p$ counter Dirac vortices through the cell, to
obtain zero net flux. Then a divergence free vector potential can be build for
instance by Fourier transform
\begin{eqnarray}\label{eq:FourierA}
B(\bbox{Q})=\frac{1}{cd}\int_{(c,d)}d^2 r \exp[-i\bbox{Q}\cdot\bbox{r}]
B(\bbox{r}), \\
A(\bbox{r})=\sum_{\bbox{Q}\neq 0}\left(\begin{array}{c}
iQ_y \\ -iQ_x \end{array}\right)\frac{\exp[i\bbox{Q}\cdot\bbox{r}]}{Q^2}
B(\bbox{Q}),\label{eq:FourierB}
\end{eqnarray}
where the sum is over  $\bbox{Q}$ in the reciprocal lattice.
Here we have used continuum notation, but it is straightforward to write
down the lattice equivalents of the expressions.

\subsection{Lattice calculation of Hall conductivity}

We have calculated the Hall conductance of the 2DEG in the vortex field by a
numerical lattice method. This we do because the calculations then reduces to
linear algebra operations on finite size matrices, which can be implemented
in a program on a computer.
The idea is to consider an electron moving on a
discrete lattice, rather than in continuum space. We know, although we are
not going to prove it here, that in the limit where the discrete lattice
becomes fine-grained compared to all other characteristic length of the system,
the continuum theory is recovered. Here we assume that the original Bravais
lattice has square lattice symmetry, with a lattice parameter which we
call~$a$. The discrete micro lattice is then introduced as a fine-grained
square lattice inside the unit cell of the Bravais lattice. The lattice
parameter of the micro lattice we take as $a/d$, where $d$ is some
large number, in order to keep the two lattices commensurable. The condition
that we have to impose on the micro lattice, in order that it is a good
approximation to the continuum, can then be stated
\begin{equation}
a/d\ll a,\lambda_F, \lambda_s,\cdots .
\end{equation}
In the numerical calculations we have made, we have taken $d=10$.

The tight-binding calculations are made with the Hamiltonian
\begin{equation}
H = -\sum_{\bbox{ij\tau\tau'}}t_{\bbox{i}+\bbox{\tau},\bbox{j}+\bbox{\tau'}}
c^{\dag}_{\bbox{i}+\bbox{\tau}}c_{\bbox{j}+\bbox{\tau'}}.
\end{equation}
Here $\bbox{i},\bbox{j}$ are Bravais lattice vectors, and
$\bbox{\tau},\bbox{\tau}'$
are vectors indicating
the sites in the basis. The matrix elements
$t_{\bbox{i}+\bbox{\tau},\bbox{j}+\bbox{\tau'}}$ are
taken non-zero only between nearest neighbour sites.
The matrix element between
two nearest neighbour sites $\bbox{\tau}$ and $\bbox{\tau'}$
are complex variables
$t_{\bbox{\tau},\bbox{\tau}+\bbox{e}_{\mu}}=t e^{iA_{\mu}(\tau)}$ with a phase
given by the vector potential $A_{\mu}(\tau)$ residing on the link joining
the sites.
The translation invariance of the Hamiltonian can
then be stated $t_{\bbox{i}+\bbox{\tau}+\bbox{l},\bbox{j}+\bbox{\tau'}+
\bbox{l}} = t_{\bbox{i}+\bbox{\tau},\bbox{j}+\bbox{\tau'}}$, for all vectors
$\bbox{l}$ belonging to the Bravais lattice.
Let us introduce the system on which our
calculations were made as an example. Fig.~\ref{basis} shows the unit cell
with its internal structure i.e. the basis.
There are $N=d\cdot d$ sites
in the basis. The length of the links we write as $a/d$, where $a$ is the
side of the unit cell, with area $\Omega = a^2$.
The vectors $\bbox{\tau}= (\tau_1,\tau_2)\frac{a}{d},$
$\tau_1,\tau_2 =0,1,\dots d-1$
are offsets into the basis, while the vectors $\bbox{i}=(i_1,i_2)a,$
$i_1,i_2 \in \Bbb{Z}$ indicate the cells in the Bravais lattice.
The operator $c^{\dag}_{\bbox{i}+\bbox{\tau}}$, for
a given $\bbox{\tau}$, is defined on the Bravais lattice, and accordingly it
can be resolved as a Fourier integral over the Brillouin zone as
\begin{equation}
c^{\dag}_{\bbox{j}+\bbox{\tau}} = \frac{\Omega}{(2\pi)^2}\int_{BZ} d^2 q
e^{-i\bbox{q} \cdot (\bbox{j} + \bbox{\tau})}c^{\dag}_{\bbox{q},\bbox{\tau}}.
\end{equation}
Inserting this and using the translation invariance, the Hamiltonian can be
rewritten as
\begin{equation}
H = \frac{\Omega}{(2\pi)^2}\int_{BZ}d^2 k H_{\bbox{k}},
\end{equation}
where we have introduced
\begin{equation}
H_{\bbox{k}} = -\sum_{\bbox{j \tau \tau'}}t_{\bbox{\tau},\bbox{j}+\bbox{\tau'}}
e^{i\bbox{k}\cdot(\bbox{j}
+ \bbox{\tau}'-\bbox{\tau})}
c^{\dag}_{\bbox{k},\bbox{\tau}}c_{\bbox{k},\bbox{\tau'}}.
\end{equation}
It is seen that $H_{\bbox{k}}$ only mixes the $N$ states $|\bbox{k\tau}
\rangle$, i.e.\ it is a $N\times N$ matrix. The $N$ eigenvalues of
$H_{\bbox{k}}$ are the energies of
the $N$ tight binding Bloch states with wave vector $\bbox{k}$. Let us denote
the eigenstates of $H_{\bbox{k}}$ by $u^{\alpha}_{\bbox{k}}$
\begin{equation}\label{eq:schrodinger}
H_{\bbox{k}} u^{\alpha}_{\bbox{k}} = E^{\alpha}_{\bbox{k}}
u^{\alpha}_{\bbox{k}}
\end{equation}
where $\alpha=1,2,\dots N$ and $E^{\alpha}_{\bbox{k}} \le
E^{\alpha +1}_{\bbox{k}}$.
{}From the $N$ dimensional vector $u^{\alpha}_{\bbox{k}}$ we can construct the
eigenstate $\Psi^{\alpha}_{\bbox{k}}$ of the Hamiltonian $H$
\begin{equation}
\left\langle \bbox{j}+\bbox{\tau} |\Psi^{\alpha}_{\bbox{k}}\right\rangle =
e^{i\bbox{k}\cdot(\bbox{j}+\bbox{\tau})} u^{\alpha}_{\bbox{k}}(\bbox{\tau}).
\end{equation}
It is straightforward to verify that this is the correct Bloch eigenstate of
$H$. The band structure can be calculated directly by diagonalizing the
$N\times N$ matrices, $H_{\bbox{k}}$, for representative choices of
$\bbox{k}$ in
the Brillouin zone. Before one can compare the spectrum obtained from this
calculation with that of a continuum system, a scaling of the energies is
required. To scale the energy to the spectrum of a particle with an effective
mass $m$, we have to take $t=\hbar^2 d^2 /ma^2$,
and $\epsilon_{\alpha}(\bbox{k})=E^{\alpha}_{\bbox{k}} + 4t$.

The Hall conductivity can be calculated by the same method as in the
homogeneous magnetic field~\cite{TKNN82,Kohmoto85}.
We have used a single particle Kubo formula to calculate the Hall
conductance
\begin{equation}
\label{eq:sigmasum}
\sigma_{xy}=\frac{i \hbar}{A_0}\sum_{E^{\alpha}<E_F <E^{\beta}}
\frac{(J_x)_{\alpha\beta}(J_y)_{\beta\alpha} - (J_y)_{\alpha\beta}
(J_x)_{\beta\alpha}}{(E^{\alpha} - E^{\beta})^2},
\end{equation}
where $J_x$,$J_y$ are the currents in the $x$,$y$ directions, and the sum is
over single particle states $|\alpha,\bbox{k}\rangle$ with energies below and
above the Fermi level $E_F$. The area of the system is denoted $A_0$.
All quantities are diagonal in $\bbox{k}$, and
therefore this index is suppressed. The summation is composed of a discrete
sum over bands, and an integral over the Brillouin zone for each band. The
Brillouin zone shown in Fig.~\ref{Brillouin} is doubly connected because the
states on the edges is to be identified according to the translation
invariance.
This gives the Brillouin zone the topology of a torus $T^2$, with
two basic non-contractible loops.
The current operator can be written
\begin{equation}
\bbox{J}=\frac{\Omega}{(2\pi)^2}\int_{BZ}d^2 k \bbox{J}_{\bbox{k}},
\end{equation}
where $\bbox{J}_{\bbox{k}} = (e/\hbar)
(\partial H_{\bbox{k}}/\partial\bbox{k})$.
By use of some simple manipulations and completeness, it is straightforward
to refrase Eq.~\ref{eq:sigmasum}
\begin{equation}\label{eq:sigma}
\sigma_{xy}=\frac{i e^2}{\hbar A_0}\sum_{E^{\alpha}<E_F}\left(
\left\langle\frac{\partial\alpha}{\partial k_x}\left|
\frac{\partial\alpha}{\partial k_y}\right.\right\rangle -
\left\langle\frac{\partial\alpha}{\partial k_y}\left|
\frac{\partial\alpha}{\partial k_x}\right.\right\rangle
\right),
\end{equation}
where $\left|\frac{\partial\beta}{\partial k_{\mu}}\right\rangle$ is
shorthand for $\frac{\partial}{\partial k_{\mu}}\left| \beta,\bbox{k}
\right\rangle$.
This formula was first derived by Thouless, Kohmoto, Nightingale and
den Nijs \cite{TKNN82},
for a noninteracting two-dimensional electron gas in a periodic scalar
potential, and a commensurate perpendicular magnetic field. It requires some
comments to be meaningful. In order to calculate $\left|\frac{\partial
\alpha}{\partial k_{\mu}}\right\rangle$ it is necessary to consider the
difference $(\left|\alpha,\bbox{k}+\delta\bbox{k_{\mu}}\right\rangle -
\left|\alpha,\bbox{k}\right\rangle)/\delta k_{\mu}$. But this difference is not
well defined as it stands, as the {\sl phase} of the states is arbitrary.
Rather than representing the state $u_{\bbox{k}}$ by a
single vector in $\Bbb{C}^{\it N}$, it should be represented by a class of
vectors
which differ one from another only by a phase. These equivalence classes are
sometimes called rays. To compare states locally, we need to project this
$U(1)$ degree of freedom out. This is done by demanding the wave function
to be real and positive, when evaluated in a fixed point,
i.e.\ $u^{\alpha}_{\bbox{k}} (\bbox{\tau}_i)=
\langle\bbox{\tau}_i |\alpha,\bbox{k}\rangle \in \Bbb{R}_{+}$.
If the wave function
happens to be zero in $\bbox{\tau}_i$, some other point
$\bbox{\tau}_j$ must be used.
When a band has a non-zero Hall conductivity, it is not possible to find
a single $\bbox{\tau}$ which work for all the states in the Brillouin zone. The
change from $\bbox{\tau}_i$ to $\bbox{\tau}_j$
which shifts the phase of the states,
is analogous to a ``gauge transformation'' on the Brillouin zone, of the set
of states. The special
combination of terms which appear in Eq.~\ref{eq:sigmasum} is gauge
invariant with respect to these special ``gauge transformations''.
If we let $|\chi^{\alpha}\rangle$ denote a state which is obtained from
$|\alpha,\bbox{k}\rangle$ by fixing the phase according to the above scheme,
the following formula for the contribution to the Hall
conductivity from a single band $\alpha$, is well defined
\begin{equation}\label{eq:sigmawelldef}
\sigma^{\alpha}_{xy}=\frac{e^2}{h}\frac{1}{2\pi i}\int_{BZ}d^2 k
\left\{\left\langle\frac{\partial\chi^{\alpha}}{\partial k_y}\left|
\frac{\partial\chi^{\alpha}}{\partial k_x}\right.\right\rangle -
\left\langle\frac{\partial\chi^{\alpha}}{\partial k_x}\left|
\frac{\partial\chi^{\alpha}}{\partial k_y}\right.\right\rangle
\right\}.
\end{equation}
  It has been shown in detail by Kohmoto \cite{Kohmoto85}, for the homogeneous
magnetic field case, that this expression is equal to minus $e^2/h$
times the first Chern
number of a principal fiber bundle over the torus.
As the first Chern number is always an integer, this has the physical
consequence that whenever the Fermi energy lies in an energy gap, the
Hall conductance is quantized.
We will use this result to
interpret certain peaks in the $\sigma_{xy}$-spectra we have calculated.
When the Fermi level is not in an energy gap of the system, we will have to
use Eq.~\ref{eq:sigma} to calculate $\sigma_{xy}$. As we shall see, in this
case there is no topological quantization of the Hall conductivity.

\subsection{Energy band crossing}\label{sec:bandcross}

In this section we study the effect on the Hall conductivity of an energy
band crossing. This has previously been discussed in different contexts by
several authors \cite{Berry84,Wilk84,Axel}.

When the shape of the magnetic field is varied, controlled by some outer
parameter $\xi$, it will happen for certain parameter values $\xi_0$, that two
bands cross, see Fig.~\ref{bandcross}.
This is the consequence of the
Wigner-von Neumann theorem, which states that three parameters are required in
the Hamiltonian in order to produce a degeneracy not related to symmetry.
Here the parameters are
$k_x,k_y$ and the outer parameter $\xi$, which in our calculation is the
exponential length of the
flux vortices from the superconductor. When the energy difference $E^+ - E^-$
between the two bands considered is much smaller than the energy distance to
the other bands, the Hamiltonian can be restricted to the subspace spanned by
the two states $|+,\bbox{k}^0\rangle$ and $|-,\bbox{k}^0\rangle$.
The point in the Brillouin zone where the degeneracy occur we
denote $\bbox{k}^0$. The Hamiltomian $H(\bbox{k})$ is
diagonal for $\bbox{k}=\bbox{k}^0$, and we denote the diagonal elements
respectively $E_0 + \epsilon$ and $E_0 - \epsilon$.
For small deviations of $\bbox{k}$ from $\bbox{k}^0$ the
lowest order corrections to the Hamiltonian is off-diagonal elements
$\Delta(\bbox{k})$ linear in $\bbox{k}-\bbox{k}^0$.
Without essential loss of generality we can assume that $\epsilon$ is
independent of $\bbox{k}$. Then $\epsilon$ plays the role of the outer
parameter controlling the band crossing. The Hamiltonian is then
approximated by
\begin{equation}\label{eq:apx}
H(\bbox{k})=\left(
\begin{array}{cc}
\epsilon & \Delta^* \\
\Delta & -\epsilon
\end{array}
\right) + E_0.
\end{equation}
The off-diagonal element is expanded as
\begin{equation}
\Delta(\bbox{k})=\alpha(k_x -k^0_x) + \beta(k_y -k^0_y)
\end{equation}
where we have introduced the matrix elements
$\alpha=\frac{\partial}{\partial k_x}\langle -,\bbox{k}^0|
H_{k}|+,\bbox{k}^0\rangle$, and
$\beta=\frac{\partial}{\partial k_y}\langle -,\bbox{k}^0|
H_{k}|+,\bbox{k}^0\rangle$.

We want to find the consequences of the energy band degeneracy, on the
topological Hall quantum numbers of the bands.
Let us define
\begin{equation}
B_{\pm}(\bbox{k})=
\left\{\left\langle\frac{\partial\pm}{\partial k_y}\left|
\frac{\partial\pm}{\partial k_x}\right.\right\rangle -
\left\langle\frac{\partial\pm}{\partial k_x}\left|
\frac{\partial\pm}{\partial k_y}\right.\right\rangle
\right\}.
\end{equation}
Then the  interesting quantities are
the integrals of $B_{+}(\bbox{k})$ and $B_{-}(\bbox{k})$, around a small
neighbourhood of the degeneracy point $\bbox{k}^0$.
It turns out, that it is the two numbers $\alpha$ and $\beta$ that control
what happens.

In general $\alpha$ and $\beta$ will be nonzero complex numbers --- nonzero
because we have assumed the degeneracy to be of first order. Let us first
consider the degenerate case where $\alpha$ and $\beta$ are linearly
dependent, i.e.\ $\alpha/\beta$ is real, or otherwise stated
Im$(\alpha^* \beta)=0$. Then by a linear transformation we can write the
Hamiltonian
\begin{equation}
H(\bbox{\kappa})=(\kappa_1+\kappa_2) \sigma^1 + \epsilon\sigma^3 =
\left(
\begin{array}{cc}
\epsilon  &  \kappa_1 +\kappa_2  \\
\kappa_1 + \kappa_2 & - \epsilon
\end{array}\right),
\end{equation}
where $\kappa_1,\kappa_2$ are rescaled momentum variables, with dimension
of energy.
The $\sigma^{\mu}$'s refers to the Pauli matrices.
The eigenvectors of this Hamiltonian can be chosen real,
and this will clearly make $B_{\pm}(\bbox{\kappa})$ vanish. In
this case we therefore conclude that there is no exchange of topological
charge. Here the word topological charge is used to denote Hall quanta.

Let us now treat the general case where $\alpha$ and $\beta$ are linearly
independent, i.e.\ Im$(\alpha^* \beta)\neq 0$. In this case we can, by a
linear transformation, write
$\Delta(\bbox{k})=\kappa_1 + i\kappa_2=\kappa e^{i\theta}$,
which defines the scaled  variables $\kappa,\theta$.
This reduces the Hamiltonian to the form
\begin{equation}
H(\kappa,\theta)=\kappa_1 \sigma^1 + \kappa_2 \sigma^2 + \epsilon\sigma^3 =
\left(
\begin{array}{cc}
\epsilon  &  \kappa e^{-i\theta} \\
\kappa e^{i\theta} & -\epsilon
\end{array}\right).
\end{equation}
Let us define $\lambda=\sqrt{\epsilon^2 + \kappa^2}$. Then the eigenvalues
of $H(\kappa,\theta)$ are $\pm\lambda$ and the two corresponding
eigenstates are
\begin{equation}
\left|\pm,\bbox{k}\right\rangle=\frac{1}{\sqrt{2}}
\left(\begin{array}{c}
\pm\sqrt{1\pm \epsilon/\lambda} \\ e^{i\theta}\sqrt{1\mp \epsilon/\lambda}
\end{array}\right).
\end{equation}
In order to calculate the integrals of the $B_{\pm}$-functions we need
to express them in terms of the $\kappa,\theta$-variables. The Jacobian of the
transformation is given by the expression
\begin{equation}
dk_x dk_y = \frac{\kappa\, d\kappa d\theta}{|\alpha_r\beta_i -
\alpha_i\beta_r|},
\end{equation}
and
\begin{eqnarray}
B_{\pm}(\kappa,\theta)& = &\frac{\alpha_r\beta_i - \alpha_i\beta_r}{\kappa}
\left\{\left\langle\frac{\partial\pm}{\partial \theta}\left|
\frac{\partial\pm}{\partial \kappa}\right.\right\rangle -
\left\langle\frac{\partial\pm}{\partial \kappa}\left|
\frac{\partial\pm}{\partial \theta}\right.\right\rangle
\right\} \nonumber \\
& = & \pm (\alpha_r\beta_i - \alpha_i\beta_r)\frac{i\epsilon}{2\lambda^3},
\end{eqnarray}
where the indices $r,i$ refer to the real and imaginary parts respectively.
We can now calculate the contribution to the Hall conductivity from each of
the bands, from the area around $\bbox{k}^0$ given by
$|\kappa |<\kappa_c$, where $\kappa_c$ is some local
cutoff parameter which limit the integration to the area where the
approximation leading to the Hamiltonian Eq.~\ref{eq:apx} is valid
\begin{eqnarray}
\Delta\sigma^{\pm}_{xy} & = & \frac{e^2}{h}\frac{1}{2\pi i}
\int dk_x \int dk_y B_{\pm}(\bbox{k}) \nonumber \\
& = & \pm\frac{e^2}{h}\frac{1}{2\pi i}\text{sign}[\text{Im} (\alpha^* \beta)]
\int_0^{\kappa_c} d\kappa \int_0^{2\pi} d\theta
\frac{i\kappa\epsilon}{2\lambda^3} \nonumber \\
& = & \pm\frac{e^2}{2h}\text{sign}[\alpha^* \beta ]
\frac{\epsilon}{|\epsilon|}
\int_0^{\kappa_c/|\epsilon|}\frac{u\,du}{(1+u^2)^{3/2}} \nonumber \\
& = & \pm\frac{e^2}{2h}
\text{sign}[\text{Im} (\alpha^* \beta)] \text{sign}[\epsilon],
\end{eqnarray}
where the last equality sign is valid when
$\kappa_c/|\epsilon|\gg 1$, i.e.\ close to the crossing where
$\epsilon=0$.
Here the factor sign$[\epsilon]$ signals that the two bands exchange
exactly one topological conductivity
quantum $e^2/h$, at the crossing. This is not surprising, because
we know that the total contribution from the states in a single band is
always an integer times the quantum $e^2/h$. Moreover it is readily
seen that in the hypothetical situation of a $n$'th order degeneracy,
i.e.\ one for which $\Delta=\kappa^n e^{i n \theta}$, $n$ quanta are
exchanged.
The total topological charge in the
band structure is conserved. The topological charge can flow around and
rearrange itself inside a band, but only be exchanged between bands in lumps
equal to an integer multiplum of the conductivity quantum
$e^2/h$. When we gradually shrink the radius of the flux vortices to
zero, the Hall effect has to disappear.
There are two mechanisms with which the Hall
effect can be eliminated. The first is by moving the topological charge up
through the band structure by exchanging quanta, resulting in a net upward
current of topological charge, eventually moving the charge up above the Fermi
surface,
where it has no effect. The second mechanism is by rearranging the topological
charge inside the bands, so that each band has a large negative charge in the
bottom, and a large positive charge in the top, but arranged in such a clever
way that charge neutrality is more or less retained for all energies. This
second mechanism will also give a net displacement of topological charge up
above the Fermi energy, because in general the Fermi surface cuts a great many
bands, and for all these bands the large negative charge, which they have
in their bottom part, will be uncompensated. To use the language of
electricity theory, we
can say that every band gets extremely polarized, resulting in a net upward
displacement current, in analogy with the situation in a strongly polarized
dielectric. Our numerical calculations indicate, that it is the second
mechanism which is responsible for the elimination of the Hall effect,
as the radius of the vortices shrinks to zero.

When two bands are nearly degenerate for some $\bbox{k}^0$,
each of the bands have concentrated  half a quantum in a small area in k-space
around $\bbox{k}^0$, and in general the topological charge piles up across
local and global gaps in the energy spectrum.
This is the reason for the oscillatory and spiky behavior
of the Hall conductance as a function of electron density, that is seen on the
calculated spectra below. It is also the reason why the numerical integration
involved in the actual evaluation of the Hall conductivity, is more tricky
than one could wish.

\subsection{Numerical results}\label{sec:num}
\subsubsection{Transverse Conductivity}

We have calculated the transverse conductivity $\sigma_{xy}$ as a function of
the integrated density of states, for electrons in a square lattice of flux
vortices, for a series of varying cross sectional shapes of the
flux vortices.
Each of the field configurations consists of a square lattice
of flux vortices with a given exponential length~$\lambda_s$. The parameter
which vary from calculation to calculation, is the dimensionless
ratio $\xi=\lambda_s /a$, where $a$ is the length of the edges of the
quadratic unit cell. The unit cell is shown
on Fig.~\ref{basis} and contain, as we have already discussed, two vortices
and a counter Dirac vortex. In
order to do the tight binding calculation a micro lattice is introduced in the
unit cell. In all the numerical calculations we present,
the micro lattice is $10\times 10$. This gives 100 energy bands
distributed symmetrically about the center on an energy scale. Out of these
only the lower part, say band 1 to 20, approximate the
real energy bands well, while the rest is significantly affected by the finite
size of the micro lattice. A careful examination of the vector potential
reveal that the symmetry of the Hamiltonian is very high for the particular
choice of unit cell shown in Fig.~\ref{basis}. The field from a single vortex
we have taken as $B_0 e^{-(|\tau_x |+|\tau_y |)/\lambda_s}$ instead of the
more realistic $B_0 e^{-|\bbox{\tau}|/\lambda_s}$. With this choice the
vector potential can be written down analytically in closed form.
This makes the calculations
simpler, and does not break any symmetry that is not already broken by the
introduction of the micro lattice. (We have made calculations of
the band structure, with both kinds of flux vortices, and the differences are
indeed very small).
The energy spectrum is invariant under
the changes $(k_x,k_y)\longmapsto (\pm k_x,\pm k_y),(\pm k_y,\pm k_x)$. This
fact is exploited to present the band structures in an economic way. The
labels $\Gamma,X$ and $M$ correspond to the indicated points in the Brillouin
zone, Fig.~\ref{Brillouin}.

In Fig.~\ref{fig:halfbands} we have plotted a selection of typical
band structures which illustrates the crossover from the completely flat
Landau bands in the homogeneous magnetic field $\xi=\infty$, to the
band structure of electrons in a square lattice of Aharonov-Bohm
scatterers with $\alpha =1/2$ at $\xi=0$.
The band structures have been found by direct
numerical diagonalization of the Hamiltonian.
(See also Fig.~\ref{fig:onebands}).

In Fig.~\ref{sigmaXY} some typical results of the numerical calculations of
the Hall conductivity are shown.
In general we have no reason to expect, that the Hall
conductivity should be isotropic as a function of the angle between the
current and the flux vortex lattice. The results we present is for a current
running along the diagonal of the square lattice, i.e.\ along the $x$-axis
in Fig.~\ref{basis}. It is seen, that whenever there is a gap in the
spectrum, the Hall conductivity gives the quantized value in agreement with
the discussion in the last section. At Fermi energies not lying in a gap
$\sigma_{xy}$ always tend to be lower than the value it has in the
homogeneous field. And in the limit $\xi \longmapsto 0$, $\sigma_{xy}$
vanishes altogether. In this limit the electron sees a periodic array of
Aharonov-Bohm scatterers, each carrying half a flux quantum, and there is no
preferential scattering to either side. We observe that when the flat Landau
bands starts to get dispersion, the contribution to the Hall effect is no
longer distributed equally in the Brillouin zone. Instead it piles up across
local and global gaps in the spectrum resulting in the spiky $\sigma_{xy}$
spectra. The density of topological charge is plotted as a function of the
filling fraction in Fig.~\ref{densit}. It is seen that for $\xi\rightarrow 0$
the distribution gets strongly polarized, with a negative contribution to
the Hall effect at the bottom part of the bands, and a positive contribution
at the uppermost part of the bands.

In Fig.~\ref{sigmaXY}, the filling fraction is
limited to values below 10, but the same behavior continues for larger
values of the filling fraction.
We have observed the spiky behavior up to $\nu = 30$, and have no reason to
believe that it should not continue to larger values.
The limitations on our calculations, comes from the fact, that we are
only able to handle matrices of limited size in the numerical calculations.
Filling fractions of order 30 correspond in this context,
to very low density electron gases, and
consequently our calculations belong to the ``quantum'' regime, i.e.\ to the
regime where $\lambda_F\gg \lambda_s$. According to the discussion in
Sec.~\ref{sec:single}, we expect a cross-over to a semiclassical regime,
for electron gases with higher density,  where
$\lambda_F\ll \lambda_s$, with a qualitatively different behavior.

It is an important question whether it is possible to observe
these features of the Hall conductivity in experiments.
The conditions, which are necessary, are that the mean free path
$l$ is long compared to all other lengths, and that $a,\lambda_s<\lambda_F$.
The last condition indicates that we are concerned with quantum magneto
transport,
in the sense that the vector potential is included in the proper quantum
treatment of the electrons. This is in contrast to the case
$\lambda_F \ll a,\lambda_s$ where the electron transport can be treated as the
semiclassical motion of localized wave packets in a slowly varying magnetic
field.
Let us assume, in order to make some estimates, that the superconductor
has a London length somewhat less than 100nm, resulting in an exponential
length of the vortices $\lambda_s$ of about 100nm at the 2DEG, after the
broadening due to the distance between the superconductor and the 2DEG has
been taken into account. In order to have a variation in the
magnetic field we should have $a>\lambda_s$, and to be in the quantum regime
$\lambda_F>a,\lambda_s$.
This gives the estimate for the electron density
$n_e \sim 10^{10}\text{cm}^{-2}$, which
is not unrealistic. The effect of the impurities is (to first order), to give
the electrons a finite lifetime. This gives a finite longitudinal conductivity
$\sigma_{xx}$ and broadens the density of states. It also introduces
localized states at the band edges (Lifshitz tails). If the field is nearly
homogeneous, we have the standard quantum Hall picture with mobility edges
above and below every Landau band, resulting in the formation of plateaus in
$\sigma_{xy}$, which only can be observed at much higher magnetic
fields, of order $10^5$G, where the filling fraction is of order one.
On the other hand when the amount of impurities is low, that
is $k_F l \gg 1$, all these effects will be small, and we expect that
the features of $\sigma_{xy}$, shown in Fig.~\ref{sigmaXY}
will have observable consequences in
$\rho_{xy}=\sigma_{xy}/(\sigma_{xx}^2 + \sigma_{xy}^2)$.

In Fig.~\ref{fig:onebands} we have plotted the band structure of electrons in
a square lattice of vortices carrying one flux quantum each. The calculation
has been made with the same basis as the band structures shown in
Fig.~\ref{fig:halfbands}, the only difference is that all fluxes have been
multiplied by a factor of 2, as can be seen from the spacing between the
Landau bands. In the limit where $\xi\rightarrow 0$,
and the vortex lattice becomes a regular array of Aharonov-Bohm scatterers,
we recover the well known band structure of free electrons. This is in
agreement with our discussion of the AB-vortex in Sec.~\ref{sec:single}.

\subsubsection{Exchange of topological quanta}

An example of exchange of topological quanta between neighbouring bands is
shown in Fig.~\ref{exchange}.
The figure is an enlargement of the band structure around the $X$ point,
showing an accidental degeneracy between the 3'ed and 4'th band,
which occur about $\xi =0.035$. Also
indicated is the Hall conductance of each band in units of $e^2/h$,
found by numerical integration. At the degeneracy it is not possible to
define the Hall conductance for the individual bands. On the Brillouin zone
torus there are two $X$
points $X_1$, $X_2$, and the bands have a first order degeneracy in
each. The numerical integration shows that two topological quanta are
transfered from the lower to the upper band,
and this is in full agreement with the discussion of Sec.~\ref{sec:bandcross}.
Exchange of topological quanta between bands is a common phenomena as $\xi$
is varyed, and this particular example has only been chosen as an illustration
of the general phenomena.

\subsubsection{Transverse conductivity in a disordered vortex phase}

When the vortices come from a thin film of superconducting material,
which have many impurities and crystal lattice defects acting as pinning
centers for the vortices,
the distribution of vortices will be disordered rather than forming a regular
Abrikosov lattice. We expect the effect of the disorder to be,
to wash out the distinctive features of the band structure, i.e.\ to
average out the characteristic fluctuations in $\sigma_{xy}$, leaving a smooth
curve in Fig.~\ref{sigmaXY} with a characteristic dimensionless
proportionality constant $s(\xi)$,
in the form $\sigma_{xy}=(e^2/h)s(\xi)\nu$, where $\nu$
is the number of electrons in the magnetic unit cell
$\nu=n_e a^2=n_e \phi_0/B$ (the filling fraction). With this conjecture
we can estimate the normalized Hall conductivity $s(\xi)$ by making a linear
fit to the calculated $\sigma_{xy}(\nu)$ distribution. In the experimental
situation $\lambda_s$ is constant, and this makes $s(\xi)$ a function
of the applied magnetic field through the relationship $\xi=\lambda_s/a
=\lambda_s\sqrt{B/\phi_0}$. In Fig.~\ref{fig:S} we have plotted $s(B)$
for a vortex exponential length $\lambda_s=80$nm.
The $s(B)$-curve shows essentially that the Hall effect of a
dilute distribution of vortices is strongly suppressed compared to the Hall
effect of a homogeneous magnetic field with the same average strength. This is
in good qualitative agreement with what is seen in the experiments of Geim et
al.~\cite{Geim92}.
When doing experiments, one is not directly measuring the conductivities,
but rather the resistivities $\rho_{xx}$, $\rho_{xy}$.
The experiments of Geim et al.\ cover the parameter range from
$\lambda_F\ll\lambda_s$ at high 2DEG densities, down to the  value
$\lambda_F /\lambda_s =0.7$ for the 2DEG with the lowest density
experimentally obtainable, where the new phenomena begin to occur.
Our numerical calculations belongs to the other side of this
cross-over where $\lambda_F \gg\lambda_s$.
The physical picture of this cross-over can be stated as follows.
On the high density side the magnetic field varies slowly over the size
of an electron wave packet for electrons at the Fermi energy, with the result
that the wave packet more or less behaves as a classical particle. On the
other side of the crossover $\lambda_F\gg\lambda_s$ the magnetic field varies
rapidly over the length-scale of a wave packet, for an electron at the Fermi
energy, and this introduces new phenomena of an essential quantum character.

\subsection{Superlattice potential}\label{sec:sup}

The general picture we have outlined so far of energy bands having
dispersion, with the dispersion giving rise to a non trivial behavior of the
Hall conductivity, is not limited to the inhomogeneous magnetic field. The
dispersion could have another origin for instance a superlattice
potential~\cite{Pffan}.
To illustrate this a series of calculations have been made on a 2DEG in a
homogeneous magnetic field, and a scalar potential which we have taken as a
square lattice cosine potential. This system have commensurability
problems because of the two ``interfering'' length scales, given respectively
by the magnetic length $l_B=\sqrt{\hbar/eB}$, and the period of the
superlattice potential $a$.
To make things as simple as possible we have fixed the
period of the cosine potential $a$, and the magnetic field strength $B$,
and only varied the amplitude of the cosine potential. Furthermore the
flux density of the magnetic field is tuned so that the flux through one unit
cell of the cosine potential is exactly one flux quantum.
The cosine potential is
\begin{equation}
U(x,y)=V_0(\cos 2\pi x/a + \cos 2\pi y/a),
\end{equation}
and the magnetic field $B=\phi_0/a^2$. The dimensionless parameter
controlling the shape of the energy band structure is in this case
\begin{equation}
v=\displaystyle{\frac{V_0}{\hbar\omega_c}}.
\end{equation}
Examples of the $\sigma_H$-spectra are shown in Fig.~\ref{super}.
It is observed that although the spectra look different from the vortex
lattice spectra, they have the same spiky nature. The spikes have the same
interpretation as in the vortex lattice system. Local spikes are due to
local gaps in the spectra. That is when to bands are close to each other
for some $\bbox{k}$ vector in the Brillouin zone, the result is a pile up of
topological charge across the gap, and this gives a spike in the
$\sigma_H$-spectra when the Fermi energy is swept across the gap.
Global spikes, that is, spikes which go all the way up to the diagonal line
indicating the Landau limit, are due to global gaps in the energy spectrum,
combined with the topological quantization.

\section{Conclusion}

In Sec.~\ref{sec:single} of this paper we have considered the longitudinal
and the transverse resistivities of a 2DEG in a disordered distribution of
flux vortices, within the theoretical framework where each scattering event
is treated independently, and the electrons are non-interacting.
The general features observed in experiments are in agreement with the
results we have outlined, but we do not have perfect quantitative agreement.
The radius of the vortices, is estimated by Geim to be $r_0\simeq 100$nm,
while we find the best fit between the calculated and the measured Hall
resistivity for $r_0\simeq 50$nm.

In Sec.~\ref{sec:array} we have considered a new kind of experiment where a
2DEG is placed in a periodic magnetic field varying on a length scale
$\lambda_s$, comparable to (or less than) the Fermi wavelength $\lambda_F$
of the electrons. In this limit, where
it is necessary to include explicitly the vector potential in a quantum
treatment of the electron motion, we expect the 2DEG to exhibit new
phenomena.
We have presented numerical results for a non-interacting 2DEG without
impurities showing characteristic spikes of the Hall conductivity versus
filling fraction, which can be understood in terms of local and global
energy gaps in the spectrum.

\section*{Acknowledgements}
We would like to express our thanks to the referee who pointed out a subtle
miscomprehension on our part, in an earlier version of this manuscript.
Finally, we would like to thank Hans L. Skriver for computer time,
B. I. Halperin for drawing our attention to Ref. \cite{Axel},
A. K. Geim for showing us the experimental data prior to publication,
and K. J. Eriksen, D. H. Lee, P. E. Lindelof, A. Smith and R. Taboryski
for stimulating discussions.

\appendix

\section{Solution of the Boltzmann equation}
\label{app:boltz}

The Boltzmann equation, linearized in the external electric field, reads
\begin{equation}\label{eq:boltz}
-e\bbox{v}\cdot\bbox{E}\frac{\partial f^0}{\partial \epsilon}=
n_{\alpha}\int\frac{d^2 q}{(2\pi)^2}\left\{f_{p+q}w_{p+q\rightarrow p} -
            f_{p}w_{p\rightarrow p+q}\right\}
   -\frac{f_{p} - f^0}{\tau_{imp}}.
\end{equation}
Here $n_{\alpha}$ is the density of vortices, and $w_{k\rightarrow k'}$ is the
scattering probability for scattering on a single vortex.
The transition probabilities $w_{p\rightarrow p+q}$ are potentially
asymmetrical quantities, due to the time reversal breaking magnetic field in
the vortices. As argued by B. I. Sturman~\cite{Sturman},
the correct form of the collision integral, even in the absence of
detailed balance, is the one given in Eq.~\ref{eq:boltz}.
The electron-vortex scattering will be
elastic. In order to solve the Boltzmann equation we Fourier transform, and
write
\begin{eqnarray}
f(k,\theta) & = & \sum_{n=-\infty}^{\infty}e^{-in\theta}f_n(k) \\
w(k,\theta) & = & \sum_{n=-\infty}^{\infty}e^{-in\theta}w_n(k),
\end{eqnarray}
where $\theta$ is the angle between $\bbox{k}$ and $\bbox{E}$.
The electron-vortex collision integral is diagonal when Fourier transformed,
and we get the following equation for the $n$'th component of the
distribution function
\begin{equation}
-evE\frac{\partial f^0}{\partial\epsilon}\frac{1}{2}
\{\delta_{n,1}+\delta_{n,-1}\} =
-n_{\alpha} \{w_0 - w_n\}f_n  -\frac{f_n - f^0\delta_{n,0}}{\tau_{imp}}.
\end{equation}
The current is given by
\begin{equation}\label{eq:current}
\bbox{j}=-2e\int\frac{d^2 k}{(2\pi)^2}\bbox{v}f(k,\theta)
=-\frac{e^2\epsilon_F E}{\pi \hbar^2 n_{\alpha}}
\left\{\begin{array}{c}
\text{Re}[\displaystyle{\frac{1}{w_1 - w_0}}] \\
\text{Im}[\displaystyle{\frac{1}{w_1 - w_0}}]
\end{array}\right\},
\end{equation}
from which the conductivities can be read off. The resistivities are found by
inverting the conductivity tensor
\begin{eqnarray}
\rho_{xy} &=& \rho_H^0\frac{k_F}{\alpha}\int^{\pi}_{-\pi}
 \frac{d\theta}{2\pi}
\sin\theta w(\theta), \label{eq:rhoxy}\\
\rho_{xx} &=& \rho_H^0\frac{k_F}{\alpha}\int^{\pi}_{-\pi}
 \frac{d\theta}{2\pi}
(1+\cos\theta ) w(\theta).
\label{eq:rhoxx}
\end{eqnarray}
For a vortex with flux $\alpha\phi_0$ we have $k_F/\alpha = 2\gamma/r_0$,
with $\gamma = l_c/r_0$ and the cyclotron radius $l_c = v/\omega_c$.

\section{Integral representation of the Aharonov-Bohm wave function,
and scattering cross section}
\label{app:integral}

In the standard geometry (Fig.~\ref{fig:qgeo}) the Aharonov-Bohm wave
function~\cite{Bohm} is
\begin{equation}
\psi = \sum_m (-i)^{|m-\alpha|} J_{|m-\alpha|}(kr) e^{im\theta}.
\label{wAB}
\end{equation}
By introducing the integral representation of the Bessel functions, the sum
can be effectuated with the result
\begin{equation}
\psi = \int_{\gamma}\frac{d\tau}{2\pi}e^{-ikr\cos\tau}
\left[ \frac{e^{i\alpha\tau}}{1 - e^{i(\tau - \theta)}} -
\frac{e^{-i\alpha\tau}}{1 - e^{-i(\tau + \theta)}}\right],
\end{equation}
where the contour $\gamma$ can be deformed into the contour shown in
Fig.~\ref{fig:contour}. The integral can now be separated naturally into two
terms. The integral along the real axis consists of a principal value
integral which vanish because the integrand is odd, and a contribution from
the two poles
\begin{equation}
e^{-ikr\cos\theta +i\alpha\theta} \equiv \psi^i,
\end{equation}
which is nothing but the incoming part of the wave function. We therefore
immediately have the following expression for the scattered part of the wave
function
\begin{equation}
\psi^s \equiv -2\sin\pi\alpha \int_0^{\infty}\frac{dt}{2\pi}
e^{ikr\cosh t}\frac{e^{i\theta}\cosh\alpha t + \cosh (1-\alpha)t}
{\cos\theta + \cosh t}.
\end{equation}
At this level the splitting $\psi = \psi^i + \psi^s$ is exact. Although it
may not be apparent, the cut in $\psi^i$ for $\theta =\pm\pi$ is
accompanied by a counter cut it $\psi^s$, as is necessary in order for
$\psi$ to be smooth. In the ``wave zone'' $kr\gg 1$ it is possible to simplify
$\psi^s$ by expanding the integral asymptotically
\begin{equation}
\psi^s = -\sin\pi\alpha e^{-ikr\cos\theta + i\theta/2}
\frac{G(\sqrt{2kr}\cos\theta/2)}{G(0)},
\end{equation}
where $G$, which is related to the error function, is given by
\begin{equation}
G(x)=\int_x^{\infty}ds\, e^{i s^2}.
\end{equation}
This expression can be further simplified when $\sqrt{2kr}\cos\theta/2 \gg 1$,
i.e., asymptotically in all but the forward direction
\begin{equation}
\psi^s = -\sin\pi\alpha \frac{e^{ikr + i\theta/2 + i\pi/4}}
{\sqrt{2\pi kr}\cos\theta/2}
\equiv -F(\theta) \frac{e^{ikr+i\pi/4}}{\sqrt{r}},
\label{eq:F}
\end{equation}
leading to the scattering cross section
\begin{equation}
|F(\theta)|^2 = \frac{1}{2\pi k} \frac{\sin^2\pi\alpha}{\cos^2\theta/2}.
\end{equation}
Although this differential cross section is singular in the forward
direction, it is completely symmetric and can therefore not give rise to a
Hall effect if substituted into the Boltzmann transport equation,
Eqns.~\ref{eq:rhoxy}-\ref{eq:rhoxx}.

\section{Interference between the incoming and scattered parts of the wave
function}
\label{app:interf}

In this appendix we want to argue that it is not the ordinary scattering
cross section which should be used in  the Boltzmann equation, but
rather a redefined ``cross section'' which take into account the interference
between the scattered and incoming parts of the electron wave function.

The particle density current, of the total wave function $\psi$
\begin{equation}
\bbox{j}=\text{Re}[\psi^* \bbox{v}\psi]
\end{equation}
can naturally be separated into the terms
\begin{equation}
\bbox{j}=
\underbrace{\text{Re}\left[\psi_i^*\bbox{v}\psi_i\right]}_{\bbox{j}^i} +
\underbrace{\text{Re}\left[\psi_i^*\bbox{v}\psi_s +
\psi_s^*\bbox{v}\psi_i\right]}_{\bbox{j}^{is}}+
\underbrace{\text{Re}\left[\psi_s^*\bbox{v}\psi_s\right]}_{\bbox{j}^s}.
\end{equation}
In terms of the density current the ordinary scattering cross section
is
\begin{equation}
\frac{d\sigma}{d\theta} = \lim_{r\rightarrow\infty}
\frac{r j_r^s(r,\theta)}{j_0},
\end{equation}
where the index $r$ refers to the radial component, and $j_0$ is the incoming
current. As is well known the scattering cross section satisfies the optical
theorem
\begin{equation}
\sigma = 2\sqrt{\frac{2\pi}{k}}\text{Im}[F(\pi)],
\end{equation}
where $F(\theta)$ is defined as in Eq.~\ref{eq:F}. The optical
theorem follows directly from particle number conservation
$\bbox{\nabla} \cdot \bbox{j} + \partial\rho/\partial t=0$.
Inside a conductor there are no
collimators, or any other devises to separate $\psi^s$ from $\psi^i$ at large
distances from the scatterer, and consequently the physically correct form of
the scattering cross section in this case is
\begin{equation}
\frac{d\tilde{\sigma}}{d\theta} = \lim_{r\rightarrow\infty}
\frac{r(j_r - j_r^i)}{j_0}=\lim_{r\rightarrow\infty}
\frac{r(j_r^s + j_r^{is})}{j_0}.
\end{equation}
The equivalent of the optical theorem for this modified cross section is
\begin{equation}
\tilde{\sigma} = 0.
\end{equation}
The interference contribution $rj_r^{is}/j_0$ is only different from zero in
the forward direction $\theta = \pm\pi$, because this is the only direction
with constructive interference. For all Hamiltonians with the mirror symmetry
$\theta\rightarrow -\theta$, the difference between $\sigma$ and
$\tilde{\sigma}$ will be proportional to $\delta(\theta - \pi)$, and will
never show up in the transport coefficients
$\rho_{xx}$ and $\rho_{xy}$, because of the geometrical factors:
\begin{eqnarray}
\rho_{xx} &\propto & \int d\theta (1+\cos\theta)\frac{d\tilde{\sigma}}{d\theta}
\\
\rho_{xy} &\propto & \int d\theta \sin\theta\frac{d\tilde{\sigma}}{d\theta}=0,
\end{eqnarray}
where $1+\cos\theta\sim (\pi -\theta)^2/2$ for $\theta\sim\pi$, thus
regularizing the integral.
But if the Hamiltonian is asymmetric in $\theta$, the interference term can
have a contribution which is proportional to the {\em derivative} of a
delta function
$\delta'(\theta - \pi)$, as we will now show.
We do this by directly calculating the quantity
\begin{equation}
\lim_{r\rightarrow\infty}\frac{k}{2\pi\alpha}\frac{r}{j_0}
\int_{-\pi}^{\pi}d\theta\, \sin\theta \,  j_r^{is}(r,\theta)
\equiv \Delta^{AB}(\alpha).
\end{equation}
After a some algebra, and dropping terms which integrate to zero,
we have
\begin{equation}
\Delta^{AB}(\alpha)=\frac{kr\sin\pi\alpha}{\pi\alpha}
\int_{-\pi}^{\pi}d\theta\, \sin\theta\cos\theta\sin(1/2 -\alpha)\theta
\left[ S(\sqrt{2kr}\cos\theta/2) - C(\sqrt{2kr}\cos\theta/2)\right],
\end{equation}
Where $S,C$ are the Fresnel sine and cosine integrals~\cite{GR}.
For $kr\rightarrow\infty$ we can evaluate the integral asymptotically
\begin{equation}
\Delta^{AB}(\alpha)=\frac{4\sin\pi\alpha\cos\pi\alpha}{\pi\alpha}
\int_0^{\infty}dx\, x \left[C(x) - S(x)\right]
=\frac{\sin 2\pi\alpha}{2\pi\alpha}.
\end{equation}
This is exactly the residual value of $\zeta_{xy}$ in the Aharonov-Bohm limit
--- due to the Iordanskii force.
We conclude that, in this particular example, using the Boltzmann transport
equation with the ordinary (quantum mechanical) scattering cross section, leads
to the erroneous conclusion of a vanishing Hall resistance. In order to obtain
the correct result it is necessary to redefine the cross section to take
into account the interference between the scattered and incoming parts of
the electronic wave function.

\begin{figure}
\caption{The scattering geometry for classical scattering on an idealized
cylindrical vortex with constant magnetic field inside and no field outside.
Here the radius of the vortex has been taken as the unit of length.}
\label{fig:geo}
\end{figure}

\begin{figure}
\caption{Differential cross section (divided by $r_0$),
and classical trajectories for four
different values of the parameter $\gamma=l_c/r_0=$ 0.025, 0.40, 1.00, 2.50}
\label{fig:ccs}
\end{figure}

\begin{figure}
\caption{The Hall efficiency factor $\zeta_{xy}(\gamma)$,
and the resistance efficiency factor $\zeta_{xx}(\gamma)$
 calculated from the classical
cross section for scattering on a magnetic flux tube. The parameter
$\gamma$ is given by the cyclotron radius divided by the radius
of the flux tube $\gamma=l_c/r_0$.}
\label{fig:zclas}
\end{figure}

\begin{figure}
\caption{The geometry of the scattering situation.}
\label{fig:qgeo}
\end{figure}

\begin{figure}
\caption{The efficiency of a dilute distribution of vortices in producing
Hall effect, compared to a homogeneous magnetic field with the same average
flux density.Curves show $\zeta_{xy}$ for $\alpha=1/4,1/2,3/4,1.$}
\label{fig:zxy}
\end{figure}

\begin{figure}
\caption{Resistance efficiency of single vortex. Curves show
$2\pi\alpha\zeta_{xx}$, for $\alpha=1/4,1/2,3/4,1$.}
\label{fig:zxx}
\end{figure}

\begin{figure}
\caption{These plots of $\zeta_{xy}$ and $2\pi\alpha\zeta_{xx}$
for a vortex with
$\alpha=10$, shows
a striking structure of resonances at the values of $k r_0$ corresponding
to the Landau quantization energies.}
\label{fig:reso}
\end{figure}

\begin{figure}
\caption{(A) The unit cell with basis. The large circles indicates
the position of the Abrikosov vortices,
and the small circle indicate the position of the
Dirac vortex with the counter flux. The micro lattice shown here is
$6\times 6$, whereas all the numerical results we have presented are
obtained with a micro lattice of $10\times 10$ sites.
(B) Four concatenated unit cells,
showing the square lattice of Abrikosov vortices.}
\label{basis}
\end{figure}

\begin{figure}
\caption{The Brillouin zone with symmetry labels.}
\label{Brillouin}
\end{figure}

\begin{figure}
\caption{Schematic energy band crossing, controlled by an outer parameter
$\gamma \propto \xi - \xi_0$.}
\label{bandcross}
\end{figure}

\begin{figure}
\caption{Band structures for 2D electrons in a square lattice of Abrikosov
vortices with $\alpha=1/2$. The different plots show band structures
corresponding to various
values of the parameter $\xi$ equal to the ratio between the exponential
length of the magnetic field from a single vortex, and the lattice parameter.}
\label{fig:halfbands}
\end{figure}

\begin{figure}
\caption{Calculated Hall conductivity versus filling fraction, for various
values of the ratio $\xi =\lambda_s /a$. These calculations are made on the
same system as the band structures shown in Fig.~11, that is,
a square lattice of
Abrikosov vortices with $\alpha=1/2$. Each of the spectra are made by
calculating the total Hall conductivity $\sigma_H(\epsilon_F)$ and the
integrated density of states $\nu(\epsilon_F)$, for
2000 equidistant values of the Fermi energy $\epsilon_F$.
The $x$-axis indicates the
integrated density of states in units of filled bands. The $y$-axis indicates
the total Hall conductivity in units of the conductivity quantum $e^2/h$.
The diagonal line in the plots indicate the Hall conductivity in a
homogeneous magnetic field.}
\label{sigmaXY}
\end{figure}

\begin{figure}
\caption{The density of Hall effect, or ``topological charge'', plotted as
function of the filling fraction, for $\alpha=1/2$.}
\label{densit}
\end{figure}

\begin{figure}
\caption{Band structures for 2D electrons in a square lattice of vortices
carrying one flux quantum each, $\alpha=1$.
The different plots shows band structures corresponding to various
values of the parameter $\xi$ equal to the ratio between the exponential
length of the magnetic field from a single vortex, and the lattice parameter.
The band structures have been calculated using the basis shown in Fig.~8,
with the only difference that
here the flux through each of the vortices are $\phi_0$, and the counter
Dirac flux is $-2\phi_0$.}
\label{fig:onebands}
\end{figure}

\begin{figure}
\caption{Exchange of topological quanta. The figure shows an enlargement of
the $\alpha=1/2$ band structure around the $X$ point, where
a degeneracy between the 3'rd and 4'th band occur. (See Fig.~11).
The parameter $\gamma$ appearing in the figure
is defined as $\gamma=\xi - \xi_{0}$, with $\xi_{0}=0.035$.
The numbers give the Hall conductance of the  bands in units of
$e^2/h$, found by numerical integration.}
\label{exchange}
\end{figure}

\begin{figure}
\caption{The normalized Hall conductivity $s(B)$ for vortices of exponential
length $\lambda_s=$80nm, and $\alpha=1/2$.}
\label{fig:S}
\end{figure}

\begin{figure}
\caption{Calculated Hall conductivity versus filling fraction, for a 2DEG in
a homogeneous magnetic field, and a square lattice cosine potential, in the
special case where the magnetic flux density is exactly equal to one flux
quantum per unit cell area. The dimensionless parameter $v$ indicated in the
plots, is equal to the amplitude of the cosine potential divided by the
Landau energy $\hbar\omega_c$.
 The $x$-axis indicates the
integrated density of states in units of filled bands. The $y$-axis indicates
the total Hall conductivity in units of the conductivity quantum $e^2/h$.
The diagonal lines in the plots indicate the Hall conductivity in a
homogeneous magnetic field, without any potential.}
\label{super}
\end{figure}

\begin{figure}
\caption{The contour of integration.}
\label{fig:contour}
\end{figure}

\end{document}